\documentclass[journal,twoside,web]{IEEEtran_adapted}

\usepackage{generic}
\usepackage{amsmath,amssymb,amsfonts}
\usepackage{dblfloatfix}
\usepackage{graphicx}
\usepackage{textcomp}
\usepackage{float}
\usepackage[normalem]{ulem}
\makeatletter
\let\NAT@parse\undefined
\makeatother
\usepackage[hidelinks]{hyperref}
\usepackage{marginnote}

\let\oldcaption\caption
\renewcommand{\caption}[1]{\oldcaption{\unboldmath #1}}

\begin{document}

\bstctlcite{ieeetran:BSTcontrol}

\title{\vspace{-4mm}Reducing Line-of-block Artifacts in Cardiac Activation Maps Estimated Using ECG~Imaging: A Comparison of Source Models and \\Estimation Methods}
\author{Steffen Schuler*, Matthias Schaufelberger, Laura R. Bear, Jake A. Bergquist, Matthijs J. M. Cluitmans, Jaume Coll-Font, Önder N. Onak, Brian Zenger, Axel Loewe, Rob S. MacLeod, Dana H. Brooks, \\and Olaf Dössel\vspace{-7mm}
\thanks{Copyright (c) 2021 IEEE. Personal use of this material is permitted. However, permission to use this material for any other purposes must be obtained from the IEEE by sending an email to pubs-permissions@ieee.org.
This work was supported by the German Research Foundation (DFG) under grant DO 637/21-1 and the state of Baden-Württemberg through bwHPC. The authors are grateful for the support of the Consortium for ECG Imaging (CEI).}
\thanks{*Asterisk indicates corresponding author. S. Schuler (e-mail: publications@ibt.kit.edu), M. Schaufelberger, A. Loewe and O. Dössel are with the Institute of Biomedical Engineering, Karlsruhe Institute of Technology, Karlsruhe, Germany.
L. R. Bear is with the IHU-LIRYC, University of Bordeaux, Inserm, U1045, CRCTB, Bordeaux, France.
J. A. Bergquist, B. Zenger and R. S. MacLeod are with the SCI Institute, University of Utah, Salt Lake City, USA.
M. J. M. Cluitmans is with CARIM, Maastricht University Medical Centre, Maastricht, the Netherlands.
J. Coll-Font is with the CBI, Massachusetts General Hospital, Harvard Medical School, Boston, USA.
Ö. N. Onak is with the Institute of Applied Mathematics, METU, Ankara, Turkey.
D. H. Brooks is with the Department of Electrical \& Computer Engineering, Northeastern University, Boston, USA.}}

\maketitle

\begin{abstract}
\textit{Objective:} To investigate cardiac activation maps estimated using electrocardiographic imaging and to find methods reducing line-of-block (LoB) artifacts, while preserving real LoBs.
\textit{Methods:} Body surface potentials were computed for 137 simulated ventricular excitations. Subsequently, the inverse problem was solved to obtain extracellular potentials (EP) and transmembrane voltages (TMV). From these, activation times (AT) were estimated using four methods and compared to the ground truth. This process was evaluated with two cardiac mesh resolutions. Factors contributing to LoB artifacts were identified by analyzing the impact of spatial and temporal smoothing on the morphology of source signals.
\textit{Results:} AT estimation using a spatiotemporal derivative performed better than using a temporal derivative. Compared to deflection-based AT estimation, correlation-based methods were less prone to LoB artifacts but performed worse in identifying real LoBs. Temporal smoothing could eliminate artifacts for TMVs but not for EPs, which could be linked to their temporal morphology. TMVs led to more accurate ATs on the septum than EPs. Mesh resolution had a negligible effect on inverse reconstructions, but small distances were important for cross-correlation-based estimation of AT delays.
\textit{Conclusion:} LoB artifacts are mainly caused by the inherent spatial smoothing effect of the inverse reconstruction. Among the configurations evaluated, only deflection-based AT estimation in combination with TMVs and strong temporal smoothing can prevent LoB artifacts, while preserving real LoBs.
\textit{Significance:} Regions of slow conduction are of considerable clinical interest and LoB artifacts observed in non-invasive ATs can lead to misinterpretations. We addressed this problem by identifying factors causing such artifacts and methods to reduce them.
\end{abstract}


\section{Introduction}
\label{sec:intro}
Electrocardiographic imaging~(ECGI) combines recordings of body surface potentials (BSP) with a model of the torso to reconstruct the electrical activity within or on the surface of the heart from non-invasive measurements~\cite{cluitmans18}.
By solving an inverse problem, ECGI aims to provide electrophysiologists with a representation of the cardiac electrical activity that contains more diagnostic information than the surface electrocardiogram (ECG). Depending on the problem formulation, several electrophysiological quantities of interest can be reconstructed using ECGI or derived from the reconstructions~\cite{pullan11,pereira20}. One of the most fundamental quantities for the analysis of arrhythmias are activation times~(AT), which represent the sequence of cardiac depolarization. A common approach to non-invasive AT mapping is to formulate the inverse problem in terms of potential-based sources and then estimate ATs from the reconstructed source signals.

This study deals with the AT estimation from two well-established surface source models: extracellular potentials~(EP) and surface transmembrane voltages~(TMV). EPs can be further divided into two subtypes: 1) pericardial potentials (PP) on a closed surface around the heart including the epi- but not the endocardium, and 2) epi- and endocardial potentials~(EEP) on a joint closed surface.
While PPs are currently most often used in clinical~\cite{rudy13,graham19} and experimental~\cite{cluitmans17,bear18,bear20} studies, some researchers have adopted the use of EEPs~\cite{erem14,wang18}.
Despite theoretical differences between EPs and TMVs, only a few studies~\cite{messnarz04a,oosterom14} directly compare these two source models and their practical impact on ECGI remains unclear.

Recent work has reported significant line-of-block (LoB) artifacts in combination with otherwise overly homogeneous regions in ATs estimated from reconstructed PPs~\cite{duchateau18,bear19}. This work has raised doubts about the capability of ECGI to distinguish between real and artifactual regions of slow conduction, a topic of considerable clinical importance. Although hypothetical explanations have been put forward~\cite{jia19}, the cause for these artifacts still is a matter of debate and effective methods to avoid or reduce them need to be established.

Here, we use computer simulations of focal ventricular excitations to investigate LoB artifacts in ECGI reconstructions. Compared to clinical or experimental studies, this approach has the advantage of having a perfect ground truth and full control over the electrophysiological properties that might be causing the artifacts.
We consider four different methods for AT estimation: deflection-based and correlation-based methods in combination with either a temporal or a spatiotemporal derivative.
By comparing source models, mesh resolutions, AT estimation methods and parameters involved in AT estimation, we aim to identify configurations that circumvent artifactual LoBs but preserve real LoBs. Furthermore, we provide insights into source-model-specific and electrophysiological properties contributing to the creation of AT artifacts.

\section{Methods}
\label{sec:methods}

\subsection{Source Models and Heart Meshes}
\label{sec:srcModels}
According to the bidomain model~\cite{geselowitz83}, the following Poisson-type equation relates TMVs $V_m$ in the myocardium with EPs $\phi_e$ in the entire body:
\begin{equation}
\nabla\cdot\left((\boldsymbol{\sigma}_i+\boldsymbol{\sigma}_e)\nabla\phi_e\right) = -\nabla\cdot(\boldsymbol{\sigma}_i\nabla V_m)
\label{eq:bidomain}
\end{equation}
BSPs are given by $\phi_e$ on the body surface. $\boldsymbol{\sigma}_i$ and $\boldsymbol{\sigma}_e$ are conductivity tensors of the intra- and extracellular domain, respectively. In this model, the electrical activity of the cardiomyocytes is represented by an impressed current density $\boldsymbol{\sigma}_i\nabla V_m$, which in turn is expressed in terms of TMVs. Equation~\eqref{eq:bidomain} states that the divergence of the gradient of TMVs is the source of EPs. By assuming continuity of the right side of~\eqref{eq:bidomain} within the myocardium and equal anisotropy ratios of conductivities in both domains, it is possible to derive TMVs on only the surface of the myocardium as an equivalent source representation~\cite{yamashita85}. In contrast to TMVs inside the myocardium, these surface TMVs can be uniquely determined from BSPs (up to an additive constant and not taking into account ambiguities arising from an imperfect forward model or noisy measurements)~\cite{kalinin17}. It is worth mentioning that the equivalent double layer~(EDL) sources~\cite{oosterom02} used in \cite{oosterom14,janssen18,dam09} only differ from surface TMVs by a constant factor.

Outside the myocardium, the right side of~\eqref{eq:bidomain} becomes zero. 
Therefore, BSPs can also be obtained using Laplace's equation with Dirichlet boundary conditions for $\phi_e$ on a closed surface encompassing the myocardium. If this surface is taken close to the heart (or as the myocardial surface itself), EPs can also be seen as a source model. As the electric field decays rapidly in space, EPs are usually dominated by the electrical activity of nearby tissue. EPs do not require assumptions about anisotropy or continuity within the tissue.

\begin{figure}[t]
    \centering
    \includegraphics[width=0.9\linewidth]{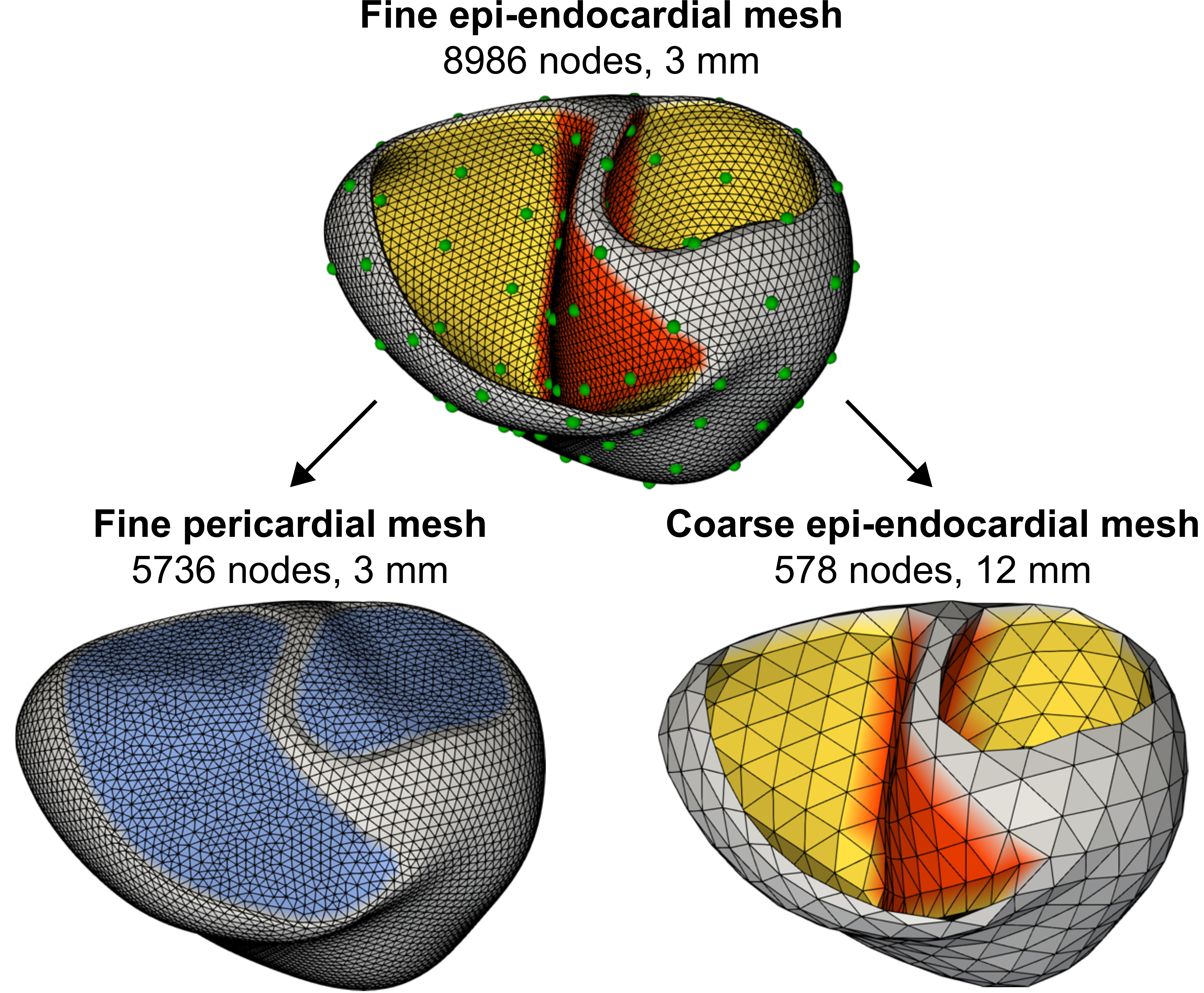}
    \caption{Surface meshes of the heart. \textit{Gray:} Epicardium. \textit{Yellow:} Non-septal endocardium. \textit{Red:} Septal endocardium. \textit{Blue:} Caps added to the epicardium to close the pericardial surface. \textit{Green spheres:} 137 uniformly distributed excitation origins. The number of nodes and mean edge length are above each panel.}
    \label{fig:hearts}
\end{figure}

We used three heart meshes in this study, shown in Fig.~\ref{fig:hearts}. The fine epi-endocardial mesh represents the baseline and was used to reconstruct EEPs and TMVs. From this mesh, a fine pericardial mesh was created by removing the endocardium (yellow and red) and closing the surface by adding caps at the base (blue). This mesh has the same resolution as the fine epi-endocardial mesh and was used to reconstruct PPs. To study the effect of mesh resolution for EEPs and TMVs, a coarse epi-endocardial mesh was created by triangulating a subset of nodes in the fine epi-endocardial mesh.

\subsection{Excitation Simulations}
\label{sec:excitSim}
\begin{figure}[t]
    \centering
    \includegraphics[width=0.95\linewidth]{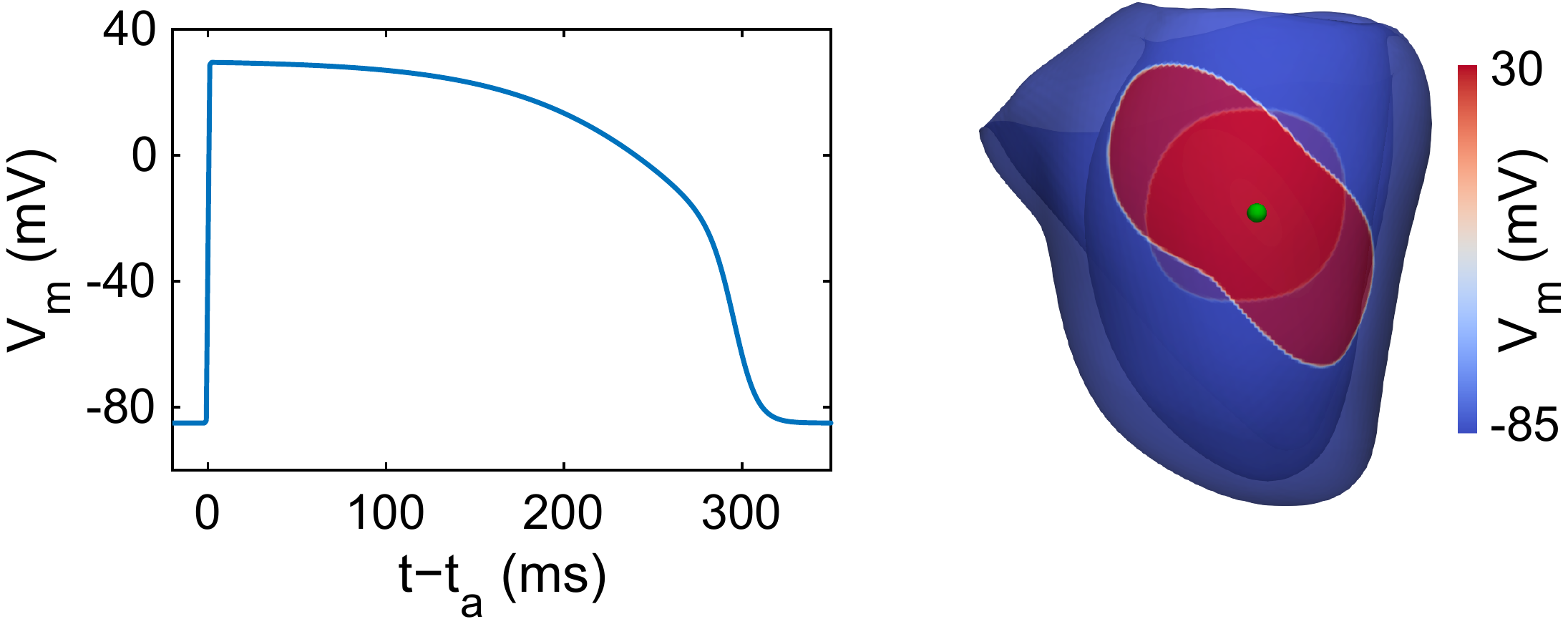}
    \caption{TMV time course (\textit{left}) and spatial distribution of TMVs 70\,ms after onset of an excitation in the lateral wall of the left ventricle (\textit{right}).}
    \label{fig:tmv}
    \vspace{-1mm}
\end{figure}

137 focal excitation simulations were performed with an eikonal model, which is able to reproduce the main characteristics of cardiac excitation spread \cite{franzone93}. This model was chosen over a monodomain model for computational speed, as a sufficiently large number of different simulation scenarios is required to come to a reliable conclusion about the general performance of an inverse method. For the eikonal simulations, the fine epi-endocardial mesh was linearly subdivided twice (mean edge length: $0.75\,\mathrm{mm}$), tetrahedralized and 137 excitation origins were uniformly distributed across the epi- and endocardium (green spheres in Fig.~\ref{fig:hearts}). Fiber orientations were assigned using the rule-based method in~\cite{bayer12}, which was adapted\footnote{\href{https://github.com/KIT-IBT/LDRB_Fibers}{https://github.com/KIT-IBT/LDRB\_Fibers}} to yield a smooth fiber field in the free walls and an approximately linear transmural fiber rotation ($\alpha_\mathrm{endo}\,{=}\,60^\circ$, $\alpha_\mathrm{epi}\,{=}\,-60^\circ$). Conduction velocities were set to $0.6\,\mathrm{m/s}$ and $0.\overline{2}\,\mathrm{m/s}$ in longitudinal and transverse fiber directions, respectively, which corresponds to an anisotropy ratio of $2.7$~\cite{kleber11}. The anisotropic eikonal equation was solved with the fast iterative method~\cite{fu13}\footnote{\href{https://github.com/KIT-IBT/FIM_Eikonal}{https://github.com/KIT-IBT/FIM\_Eikonal}}, resulting in the ground truth ATs $t_a$. The following time course~\cite{dam09} was then prescribed to obtain TMVs at a sampling rate of $1\,\mathrm{kHz}$:
\begin{gather}
    V_m(t) = V_\mathrm{rest} + A\,L(t,t_a,\beta_a)\,L(t,t_r,\beta_{r1})\,L(t,t_r,\beta_{r2})\\
    \text{with}\nonumber\\
    L(t,\tau,\beta) = \frac{1}{1+e^{-\beta(t-\tau)}},\nonumber\\
    V_\mathrm{rest}=-85\,\mathrm{mV}, A=115\,\mathrm{mV}, t_r=t_a+297\,\mathrm{ms},\nonumber\\
    \beta_a=5\,\mathrm{ms}^{-1}, \beta_{r1}=-0.152\,\mathrm{ms}^{-1}, \beta_{r2}=-0.0183\,\mathrm{ms}^{-1}\nonumber
    \label{eq:tmvTimeCourse}
\end{gather}
Fig.~\ref{fig:tmv} illustrates the time course and the spatial distribution of TMVs for a lateral excitation origin in the left ventricle (LV).

\subsection{Forward Calculation}
\label{sec:fwdCalc}
\begin{figure}[ht]
    \centering
    \includegraphics[width=\linewidth]{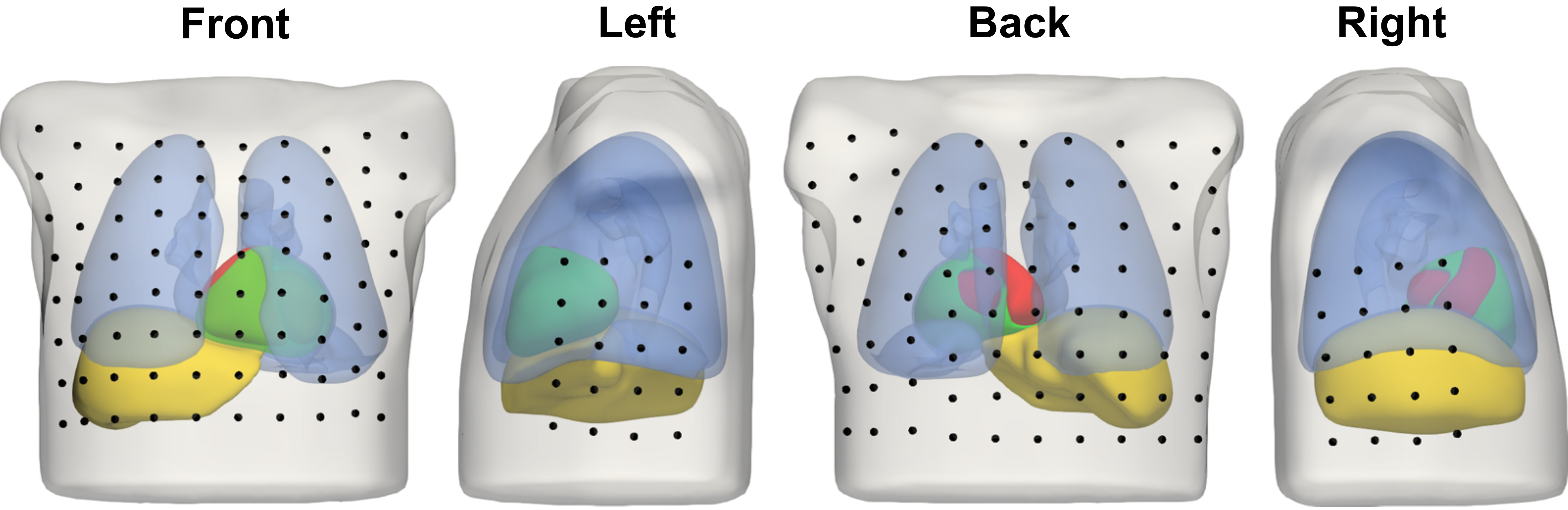}
    \caption{Inhomogeneous torso model. \textit{Green:} Myocardium (0.2\,S/m). \textit{Red:} Intracavitary blood (0.6\,S/m). \textit{Blue:} Lungs (0.04\,S/m). \textit{Yellow:} Liver (0.4\,S/m). \textit{Gray:} Rest of the torso (0.2\,S/m). \textit{Black dots:} 200 electrodes.}
    \label{fig:torso}
    \vspace{-2mm}
\end{figure}

For forward calculations, the TMVs were Laplacian downsampled \cite{schuler19} to a mesh with $1.5\,\mathrm{mm}$ mean edge length, obtained by subdividing the finer mesh in Fig.~\ref{fig:hearts} once.
The boundary element method~(BEM)~\cite{stenroos07} was then used to compute BSPs at 200 electrodes on an isotropic, inhomogeneous torso model depicted in Fig.~\ref{fig:torso}. Conductivities of the individual compartments are given in the caption. The intracellular conductivity of the myocardium was set to $0.05\,\mathrm{S/m}$. White Gaussian noise was added to the BSPs to yield a signal-to-noise ratio~(SNR) of $20\,\mathrm{dB}$ (average across all electrodes during the activation interval). This SNR is in agreement with the average value of $19.2\pm6.2\,\mathrm{dB}$ reported in~\cite[Table I]{lian02}. A different realization of noise was used for each excitation.

\subsection{Inverse Reconstruction}
\label{sec:invRec}
For inverse reconstructions, perfect knowledge of the forward model was assumed and transfer matrices for the three meshes in Fig.~\ref{fig:hearts} were obtained using BEM. Although more sophisticated, spatio-temporal regularization methods exist~\cite{greensite98,erem14,messnarz04,ghodrati06,schuler19a}, we decided to use purely spatial second-order Tikhonov regularization for this comparison study because it is widely used in combination with both considered source models~\cite{milanic14,messnarz04a}.
The inverse solution was computed for all time steps simultaneously:
\begin{equation}
    \widehat{\mathbf{X}} = \arg\min_\mathbf{X}\bigl( \|\mathbf{A}\mathbf{X}-\mathbf{B}\|_\mathrm{F}^2 + \lambda\,\|\mathbf{L}\mathbf{X}\|_\mathrm{F}^2 \bigr)
    \label{eq:tikh}
\end{equation}
Here, $\|\cdot\|_\mathrm{F}$ denotes the Frobenius norm. $\mathbf{B}\,{\in}\,\mathbb{R}^{M\times K}$ are the BSPs, $\mathbf{X}\,{\in}\,\mathbb{R}^{N\times K}$ are the cardiac sources and $\mathbf{A}\,{\in}\,\mathbb{R}^{M\times N}$ is the transfer matrix linking the two. $M$ is the number of electrodes, $N$ the number of nodes on the cardiac surface and $K$ the number of time steps. $\lambda$ is the regularization parameter and $\mathbf{L}\,{\in}\,\mathbb{R}^{N\times N}$ is a discrete Laplacian operator for the cardiac surface mesh. For PPs, we use a Laplacian acting only along the surface:
\newpage
\noindent
\begin{equation}
    \mathbf{L} = \mathbf{M}^{-1}\,\mathbf{C}
    \label{eq:Laplacian1}
\end{equation}
$\mathbf{C}$ and $\mathbf{M}$ are the cotangent stiffness and the Voronoi mass matrix of the pericardial surface mesh, respectively~\cite[Sections~2.1.2~and~2.1.3]{jacobson13}. They were computed using the gptoolbox~\cite{jacobson18}.
For formulations of ECGI including the epi- and the endocardium, transmural regularization is beneficial~\cite{erem14,schuler17}. For EEPs and TMVs, we therefore use a transmural Laplacian that acts both along and across the myocardial wall. To construct this operator, a volumetric Laplacian $\mathbf{L}_\mathrm{v}$ is first computed analogously to~\eqref{eq:Laplacian1} but for a tetrahedral mesh conforming to the epi-endocardial surface mesh. $\mathbf{L}_\mathrm{v}$ is used to obtain a matrix $\mathbf{R}_\mathrm{v\leftarrow s}$ that performs Laplacian interpolation from the surface to the volume mesh~\cite[Method~B]{oostendorp89}. Furthermore, a matrix $\mathbf{R}_\mathrm{s\leftarrow v}$ is defined that extracts the surface values from the volume mesh. The transmural Laplacian for the epi-endocardial surface mesh is then given by:
\begin{equation}
    \mathbf{L} = \mathbf{R}_\mathrm{s\leftarrow v}\,\mathbf{L}_\mathrm{v}\,\mathbf{R}_\mathrm{v\leftarrow s}
    \label{eq:Laplacian2}
\end{equation}

The L-curve criterion~\cite{hansen00} was used to determine $\lambda$. To prevent overregularization due to time steps with little activity, only the QRS interval was used to compute the L-curve, which was defined from where the spatial standard deviation of BSPs first exceeds $5\,\%$ of its maximum value to where it falls back below $25\,\%$. The resulting $\lambda$ was applied for all time steps.

After reconstruction, the baseline correction described in~\cite[Section~2.1.4]{schuler19a} was applied to TMVs. This is necessary to obtain a meaningful time course, because according to~\eqref{eq:bidomain}, only the gradient of TMVs contributes to BSPs and a different, spatially constant offset may initially be superimposed on the TMVs of each time step.

\subsection{Activation Time Estimation}
\label{sec:atEstim}
Several methods have been proposed for the estimation of ATs from cardiac mapping data~\cite{cantwell15}. We consider two types of methods that have shown to be useful for ECGI: Deflection-based~(Defl) and correlation-based~(Corr) methods.

The most basic form of Defl methods consists of locating the time of minimum or maximum temporal derivative of EPs or TMVs, respectively. This method will be called Defl-temporal~(Defl-T). In recognition of the fact that local tissue activation is not only characterized by a rapid change of sources in time but also in space, Erem et al. extended this method to use the product of the spatial and the temporal derivative as a ``spatiotemporal'' derivative~\cite{erem11,cluitmans16}. This method will be called Defl-spatiotemporal~(Defl-ST).

Correlation-based (Corr) methods~\cite{duchateau17,zhou18} use cross-correlation of the temporal or spatiotemporal derivative signal to determine time delays between nearby nodes. A spatial regression~\cite{dubois12} is then used to estimate an AT field from these delays. Depending on the type of the derivative signal, this method will be called Corr-temporal~(Corr-T) or Corr-spatiotemporal~(Corr-ST).

\begin{figure}[b]
    \centering
    \includegraphics[width=\linewidth]{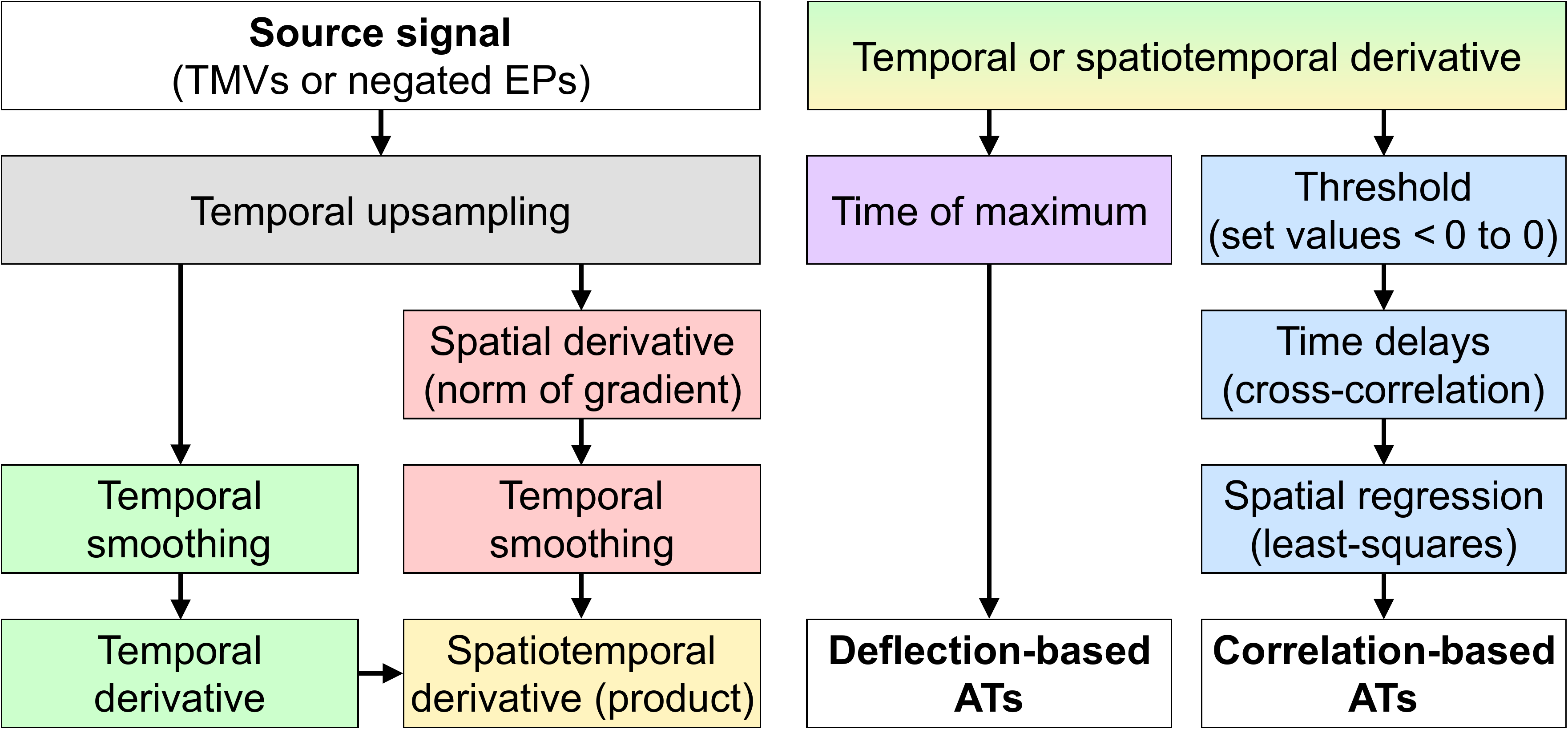}
    \caption{AT estimation pipeline. \textit{Left:} Steps to compute the temporal (green), spatial (red) and spatiotemporal (yellow) derivative. \textit{Right:} Steps to obtain ATs from the temporal or spatiotemporal derivative using the deflection-based (purple) or correlation-based (blue) method.}
    \label{fig:atEstimPipeline}
\end{figure}

Fig.~\ref{fig:atEstimPipeline} shows the AT estimation pipeline used in this study.
The key steps, as well as important numerical aspects, will be detailed in the following.

\textit{Temporal upsampling:} As the delays between nodes can be small, temporal upsampling is important for Corr methods. We found that a sufficiently large upsampling factor leads to comparably accurate delays as methods based on the Hilbert transform~\cite{shors96}. We used 10-fold temporal upsampling based on linear interpolation. For consistency, upsampling was also applied for Defl methods.

\textit{Temporal smoothing} can help to make the AT estimation less ambiguous by attenuating small, local deflections. We used a zero-phase Gaussian filter (total order: 2) with standard deviation $\sigma$ as discussed below. Border values were replicated before filtering.

\textit{Temporal derivative:} We used 3-point central differences to compute temporal derivatives. It is important to apply temporal smoothing before computing the derivative, otherwise the replication of noisy derivative values at the borders can lead to distortions in the derivative after smoothing.

\textit{Spatial derivative:} The surface gradient was computed using the gptoolbox~\cite{jacobson18} as the area-weighted average of per-face gradients around a node (method AGS in~\cite{livesu18}). No spatial smoothing is required before computing the gradient because second-order Tikhonov regularization already yields spatially smooth results. As the $L^2$ norm is a non-linear operation, it is important to apply the temporal smoothing after computation of the spatial derivative. This results in a spatial derivative with a similar spread in time as the temporal derivative.

\textit{Threshold:} As only the upslope of TMVs (or negated EPs, i.e. $-\phi_e$) corresponds to local activation, we excluded downslope portions of the signal from affecting the cross-correlation by setting negative values of the derivative to zero.

\textit{Time delays} were computed for pairs of nodes with a predefined graph distance on the surface mesh. For comparability of mesh resolutions, we will use the mean distance in $\mathrm{mm}$ instead of the number of edges to denote the ``node pair distance'' $d$.

\textit{Spatial regression:} A least-squares regression as described in~\cite{dubois12} was used to estimate the ATs from the delays.

All AT estimation methods depend on the temporal smoothing parameter~$\sigma$. To study its influence, it was varied between $5\,\mathrm{ms}$ (very mild smoothing) and $70\,\mathrm{ms}$ (strong smoothing). Corr methods additionally depend on the node pair distance~$d$. If not stated otherwise, $d$ was set to 1 mean edge length.

ATs were computed over the period from $50\,\mathrm{ms}$ before to $50\,\mathrm{ms}$ after the activation interval in order to avoid artifactual ATs being detected during the repolarization interval.

\subsection{Visualization Using Polar Projections}
\label{sec:visPolar}
\begin{figure}[H]
    \centering
    \includegraphics[width=0.8\linewidth]{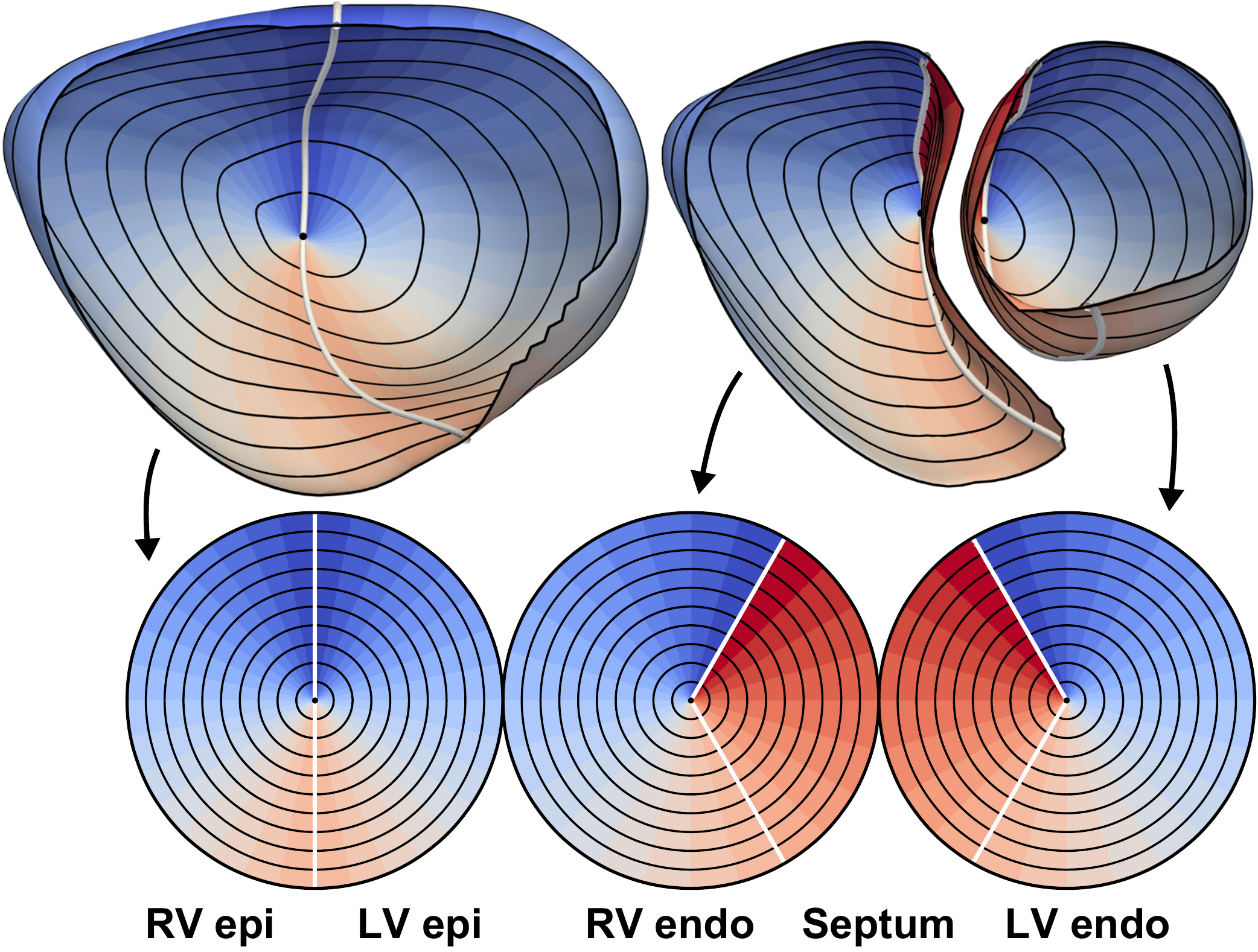}
    \caption{Rotational coordinate (colors) and apicobasal coordinate (black contours) used for polar projections of the epi- and endocardial surfaces.}
    \label{fig:coords}
\end{figure}

For a compact visualization of epi- and endocardial ATs without occlusions, we use polar projections obtained using the Consistent Biventricular Coordinates~\cite{schuler21}. Fig.~\ref{fig:coords} depicts the correspondence between the 3D surfaces and the projections.

\subsection{Metrics}
\label{sec:metrics}
\begin{figure}[ht]
    \centering
    \includegraphics[width=\linewidth]{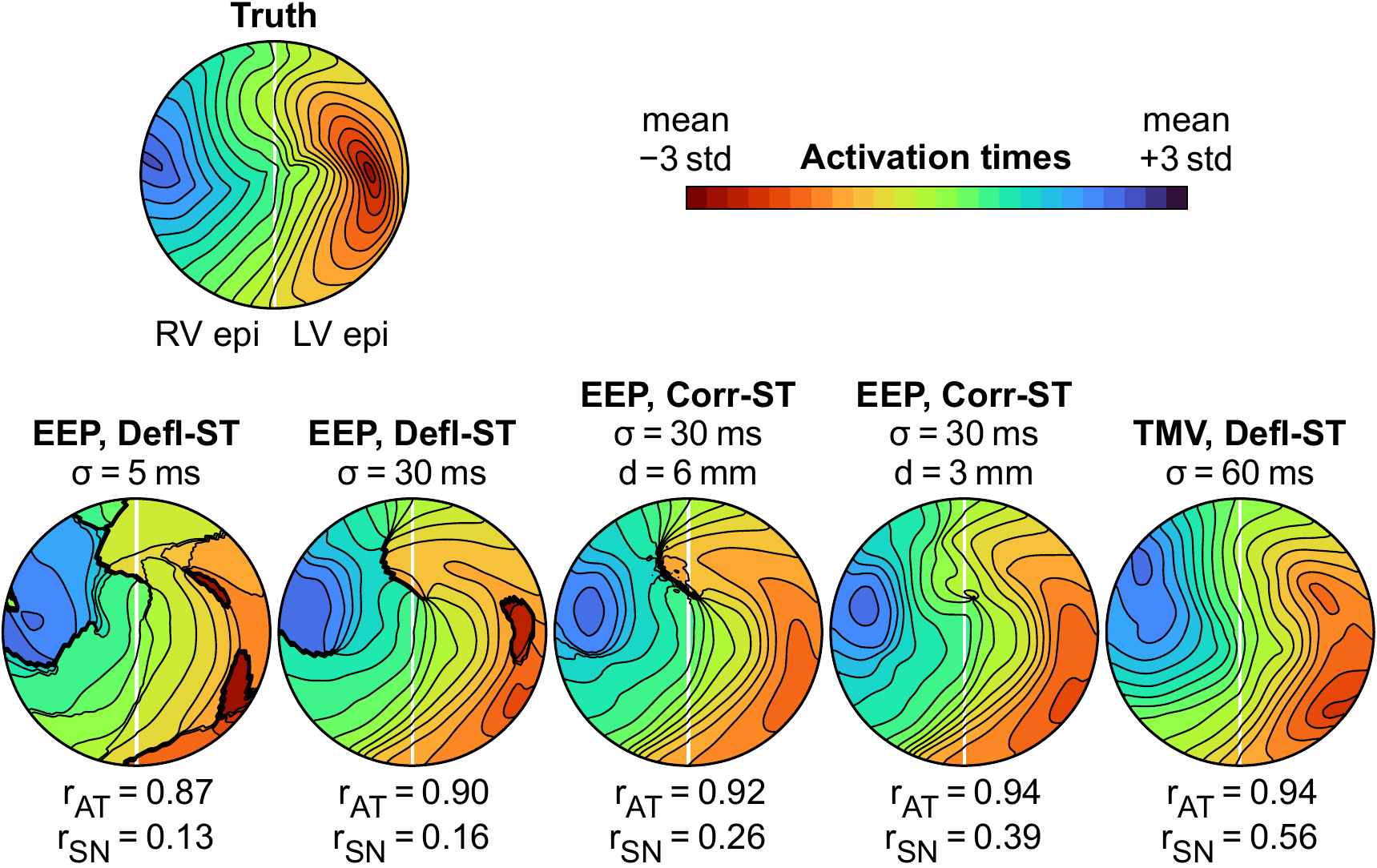}
    \caption{Metrics for 5 estimated AT maps. The slowness correlation $r_\mathrm{SN}$ is more sensitive to local artifacts than the AT correlation $r_\mathrm{AT}$.}
    \label{fig:metrics_illustration}
    \vspace{-3mm}
\end{figure}

For quantitative evaluation, we use the Pearson correlation coefficient between estimated and true ATs, denoted as $r_\mathrm{AT}$. This metric reflects the global match of two AT maps but has low sensitivity to local features like LoB artifacts. Therefore, it is complemented by a second metric: the Pearson correlation coefficient between all $3\,N$ values of the estimated and true slowness vectors, denoted as $r_\mathrm{SN}$. The slowness vector is computed as the surface gradient of ATs. The slowness was preferred to its reciprocal, the velocity, because it has well-defined values for all AT maps that might occur, whereas the velocity goes to infinity in regions of constant AT. Fig.~\ref{fig:metrics_illustration} shows both metrics for various estimated AT maps and demonstrates the better sensitivity of $r_\mathrm{SN}$ to local features compared to $r_\mathrm{AT}$.\\
The Wilcoxon signed-rank test is used to assess the significance of differences in metrics between methods ($p$-values).\\
Medians, lower and upper quartiles and $p$-values are computed across the 137 excitation origins, with each origin including a different realization of BSP noise.

\section{Results}
\label{sec:results}

\subsection{Source Models and Estimation Methods}
\label{sec:sm}
\begin{figure*}[t]
    \vspace{-2mm}
    \centering
    \includegraphics[width=\linewidth]{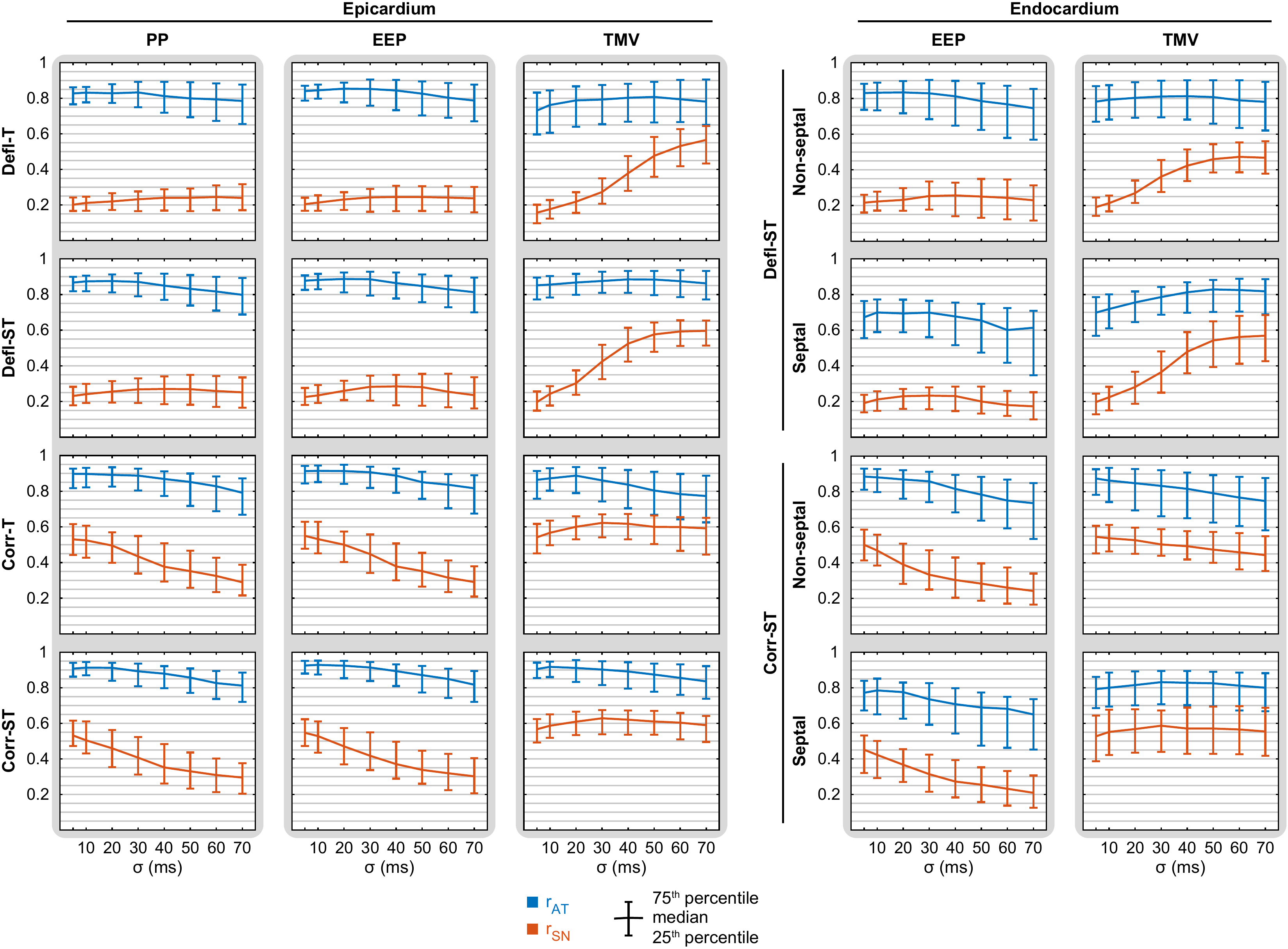}
    \caption{AT correlation $r_\mathrm{AT}$ and slowness correlation $r_\mathrm{SN}$ for ATs on the fine mesh. For each combination of source representation and AT estimation method, the dependency on the temporal smoothing parameter $\sigma$ is shown. Statistics were computed over all 137 excitations.}
    \label{fig:metrics_sm}
\end{figure*}

\begin{figure*}
    \centering
    \includegraphics[width=\linewidth]{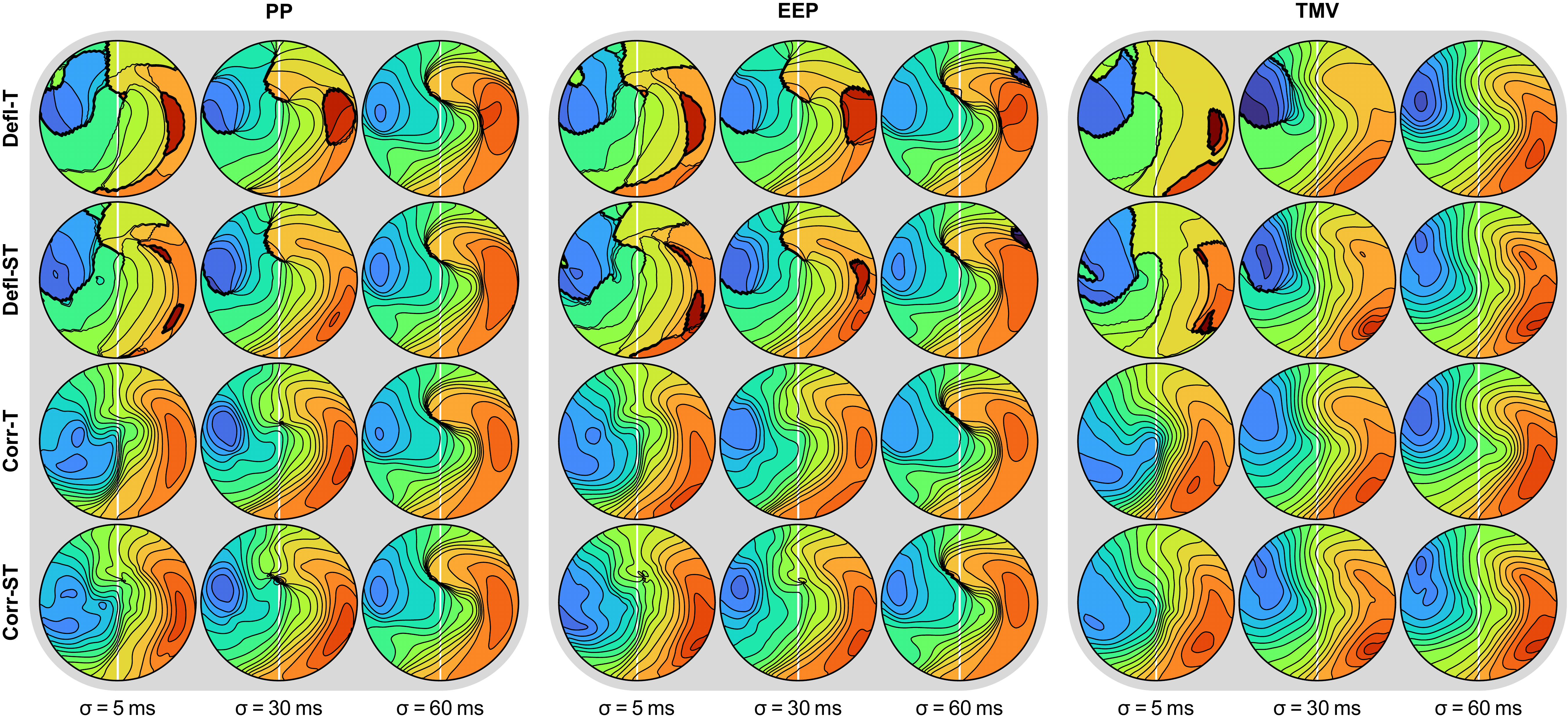}
    \caption{ATs on the epicardium of the fine mesh for the same excitation origin as shown in Fig.~\ref{fig:metrics_illustration} (see there for the ground truth and the color bar). Depicted are the results for three source representations (PP, EEP, TMV), four AT estimation methods (Defl-T, Defl-ST, Corr-T, Corr-ST) and three temporal smoothing parameters ($\sigma$).}
    \label{fig:polar_sm_epi}
\end{figure*}

We start with the comparison of source models and AT estimation methods for the fine mesh resolution and for simulations without a real LoB.
As PPs do not cover the endocardium, we first concentrate on ATs on only the epicardium (as defined in Fig.~\ref{fig:hearts}). The left panel of Fig.~\ref{fig:metrics_sm} shows the metrics for all three source model variants (columns) and all four estimation methods (rows). Each diagram contains eight levels of temporal smoothing ($\sigma$) and each bar contains 137 excitation origins. Corresponding AT maps for three representative values of $\sigma$ and an exemplary excitation origin on the lateral epicardium of the LV are visualized in Fig.~\ref{fig:polar_sm_epi}.


Epicardial ATs for both types of EPs (PPs and EEPs) were very similar, both in terms of metrics and visually, even though different Laplacian operators (non-transmural vs. transmural) were used for reconstruction.
Independent of the source model, methods using a spatiotemporal derivative~(ST) performed better than their purely temporal counterparts~(T). When we used the best $\sigma$'s tested for each case, similar levels of $r_\mathrm{AT}$ could be reached for EEPs and TMVs using both Defl-ST and Corr-ST, with Corr-ST resulting in slightly higher correlation values. In terms of $r_\mathrm{SN}$, however, Defl-ST performed substantially better for TMVs than for EEPs (median: $0.596$~vs.~$0.285$, $p\,{<}\,0.001$), indicating less artifacts in AT maps. For TMVs, $r_\mathrm{SN}$ gradually increased with $\sigma$, reaching a plateau at about $60\,\mathrm{ms}$. For EEPs, $r_\mathrm{SN}$ increased only up to about $40\,\mathrm{ms}$ but stayed below half the maximum value for TMVs and decreased again for larger values of $\sigma$. This dependency on $\sigma$ is also reflected in Fig.~\ref{fig:polar_sm_epi}. For $\sigma\,{=}\,5\,\mathrm{ms}$, the AT maps obtained with Defl methods showed large homogeneous regions and LoB artifacts (crowded isochrones), of which some appear at similar locations for EEPs and TMVs. For TMVs, the artifacts disappeared with increasing $\sigma$ and the map at $\sigma\,{=}\,60\,\mathrm{ms}$ was close to the ground truth (shown in Fig.~\ref{fig:metrics_illustration}). For EEPs, the jaggedness was also reduced with increasing $\sigma$, but distinct LoB artifacts remained or even emerged for very large values. Corr methods were also sensitive to $\sigma$, but the best results were obtained for smaller values of $\sigma$ than for Defl methods. For Corr-ST, $r_\mathrm{SN}$ showed a maximum at $\sigma\,{=}\,5\,\mathrm{ms}$ for EEPs and $30\,\mathrm{ms}$ for TMVs, with a slightly higher value for TMVs ($0.547$
~vs.~$0.628$, $p\,{<}\,0.001$). Only Corr methods in combination with a small $\sigma$ ($5\,\mathrm{ms}$) could prevent LoB artifacts for EEPs, while Defl-ST with a large $\sigma$ ($60\,\mathrm{ms}$) or Corr-ST with a medium $\sigma$ ($30\,\mathrm{ms}$) performed similarly well for TMVs.

\begin{figure*}[t]
    \centering
    \includegraphics[width=\linewidth]{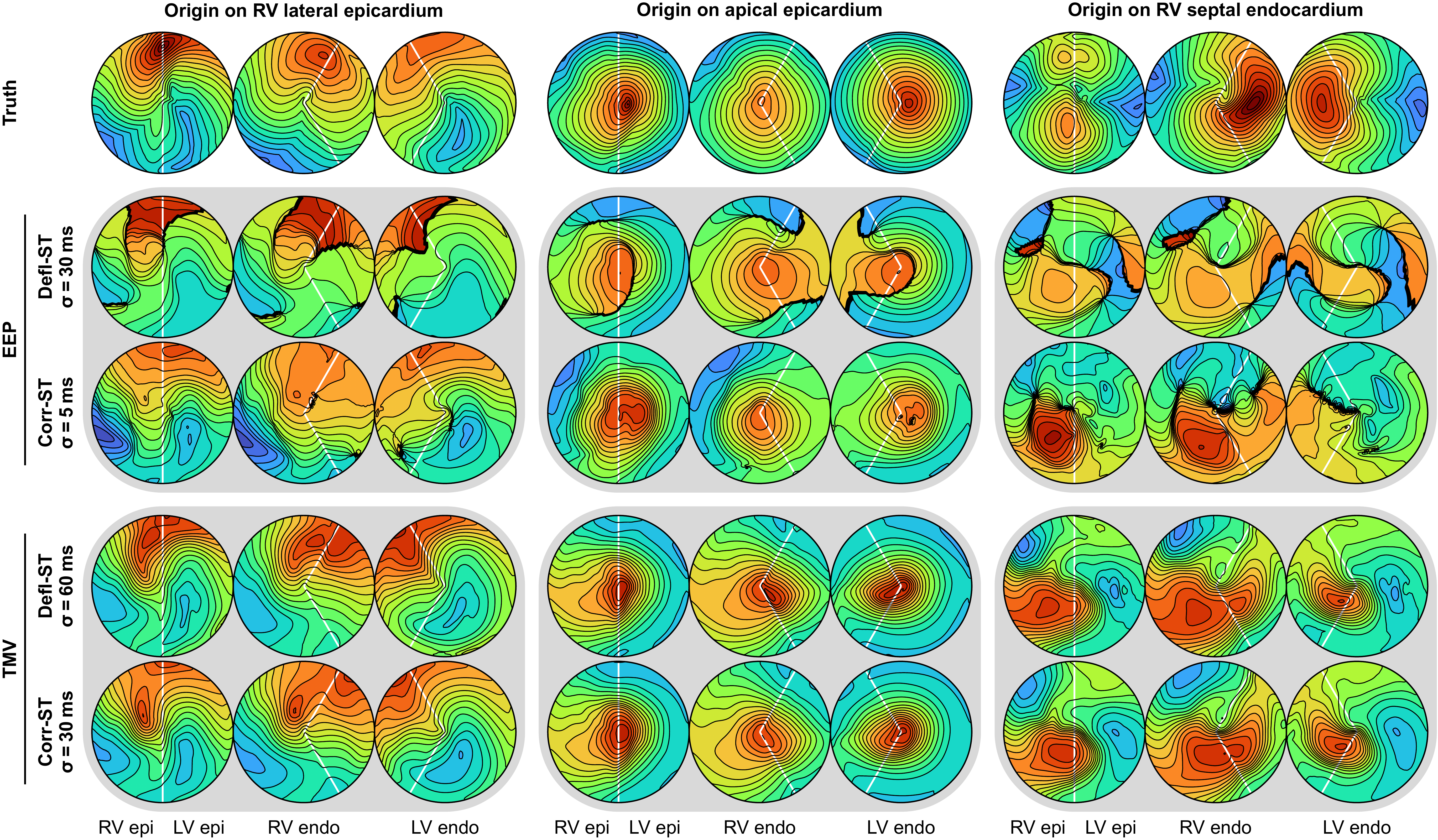}
    \caption{ATs on the epi- and endocardium of the fine mesh for three exemplary excitation origins. See Fig.~\ref{fig:metrics_illustration} for the color bar.}
    \label{fig:polar_sm_epiEndo}
\end{figure*}

As only minor differences between PPs and EEPs and consistently better results for ST methods could be found, we restricted further comparisons to only EEPs and TMVs in combination with Defl-ST and Corr-ST. The right panel of Fig.~\ref{fig:metrics_sm} shows the corresponding results on the endocardium, which is further divided into non-septal and septal regions. While $r_\mathrm{AT}$ for Defl-ST was similar for EEPs ($\sigma\,{=}\,30\,\mathrm{ms}$) and TMVs  ($\sigma\,{=}\,60\,\mathrm{ms}$) on the non-septal endocardium ($0.829$~vs.~$0.790$, $p\,{\approx}\,0.147$), TMVs showed higher values on the septal endocardium ($0.698$~vs.~$0.825$, $p\,{<}\,0.001$). A similar difference could be observed for $r_\mathrm{SN}$ on the septal endocardium, even for Corr-ST at $\sigma\,{=}\,5\,\mathrm{ms}$ and $\sigma\,{=}\,30\,\mathrm{ms}$ for EEPs and TMVs, respectively ($0.452$~vs.~$0.586$, $p\,{\approx}\,0.025$). For both source models and estimation methods, the spread of $r_\mathrm{AT}$ was higher on the endo- than on the epicardium. Fig.~\ref{fig:polar_sm_epiEndo} depicts ATs on both the epi- and the endocardium for three additional excitation origins. Here, close-to-optimal values of $\sigma$ were chosen for each configuration. It can be seen that TMVs yielded more accurate results on the septum than EEPs. This difference was most pronounced for an excitation starting on the RV septum, although the origin was wrongly indicated on the RV anterior wall for both source models.

\subsection{Mesh Resolution and Node Pair Distance}
\label{sec:res}
\begin{figure*}
    \vspace{4mm}
    \centering
    \includegraphics[width=\linewidth]{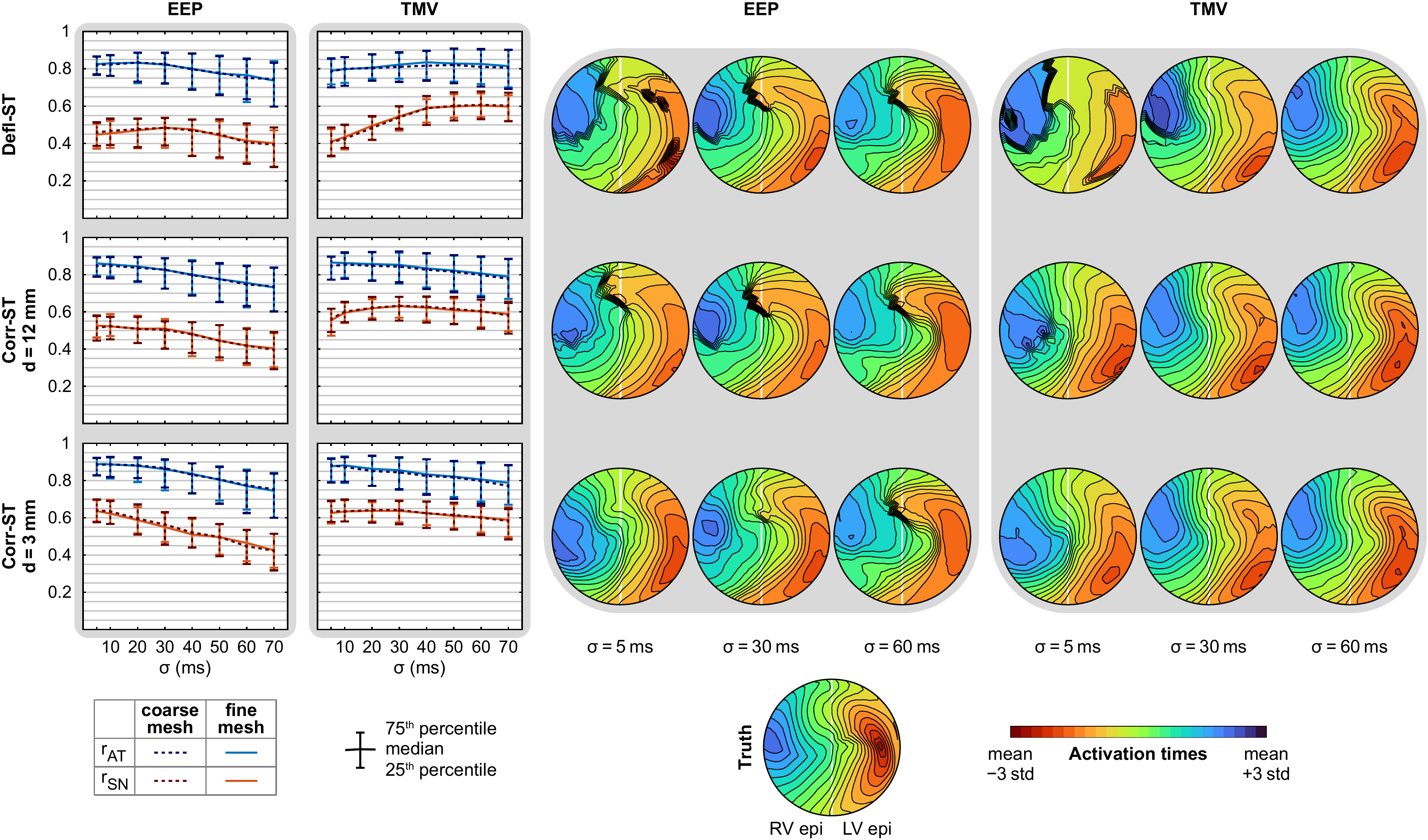}
    \caption{\textit{Left:} Comparison of metrics for ATs obtained on the coarse mesh (dark, dashed lines) and on the fine mesh (bright, solid lines). \textit{Right:} Visualization of ATs obtained on the coarse mesh (compare with the corresponding cases in Fig.~\ref{fig:polar_sm_epi}).}
    \label{fig:metrics_polar_res}
\end{figure*}

The mesh resolution changes the number of unknowns in the inverse problem, the numerical approximation of the Laplacian used for regularization, and the numerical approximation of the gradient used for AT estimation with ST methods. However, the effective impact on the resulting ATs is unclear.
Therefore, we additionally computed results using the coarse mesh and compared them to the results for the fine mesh (Fig.~\ref{fig:metrics_polar_res}). As the same resolution of ground truth must be used to allow a comparability of metrics, ATs for the fine mesh were extracted at nodes also available on the coarse mesh and metrics for both resolutions were computed on the coarse mesh.

The results for Defl-ST are given in the first row of Fig.~\ref{fig:metrics_polar_res}. The metrics for the coarse mesh (dark, dashed lines) and for the fine mesh (bright, solid lines) were almost identical. A comparison of AT maps obtained on the coarse mesh (polar projections next to the metrics) with the corresponding cases in Fig.~\ref{fig:polar_sm_epi} also revealed strong similarity. LoB artifacts at similar locations occurred for EEPs or small values of $\sigma$, although they may appear less severe due to the inherently smaller maximum gradient that can be represented on coarser meshes (this also explains the larger minimum $r_\mathrm{SN}$ compared to Fig.~\ref{fig:metrics_sm}).

The second row of Fig.~\ref{fig:metrics_polar_res} shows the results for Corr-ST and a node pair distance of $d\,{=}\,12\,\mathrm{mm}$. This is the minimal distance directly available on the coarse mesh (one edge length).
AT maps for this configuration also showed LoB artifacts for EEPs, although none could be observed for the fine mesh in the previous section. This difference could be traced back to the smaller node pair distance used as default on the fine mesh ($d\,{=}\,3\,\mathrm{mm}$, one edge length). Computing the time delays and the spatial regression for pairs of nodes with the same physical distance ($d\,{=}\,12\,\mathrm{mm}$, four edge lengths) also led to LoB artifacts for the fine mesh and resulted in similar metrics as for the coarse mesh.

To see whether cross-correlation across shorter distances could also help for the coarse mesh, we applied Laplacian interpolation of the sources onto a two-fold subdivided version of the coarse mesh. This makes it possible to use $d\,{=}\,3\,\mathrm{mm}$ also for the coarse mesh. The results are presented in the last row of Fig.~\ref{fig:metrics_polar_res} and reveal that this procedure is equally suitable for avoiding LoB artifacts as using a fine mesh directly for inverse reconstructions.

\begin{figure}[ht]
    \centering
    \includegraphics[width=\linewidth]{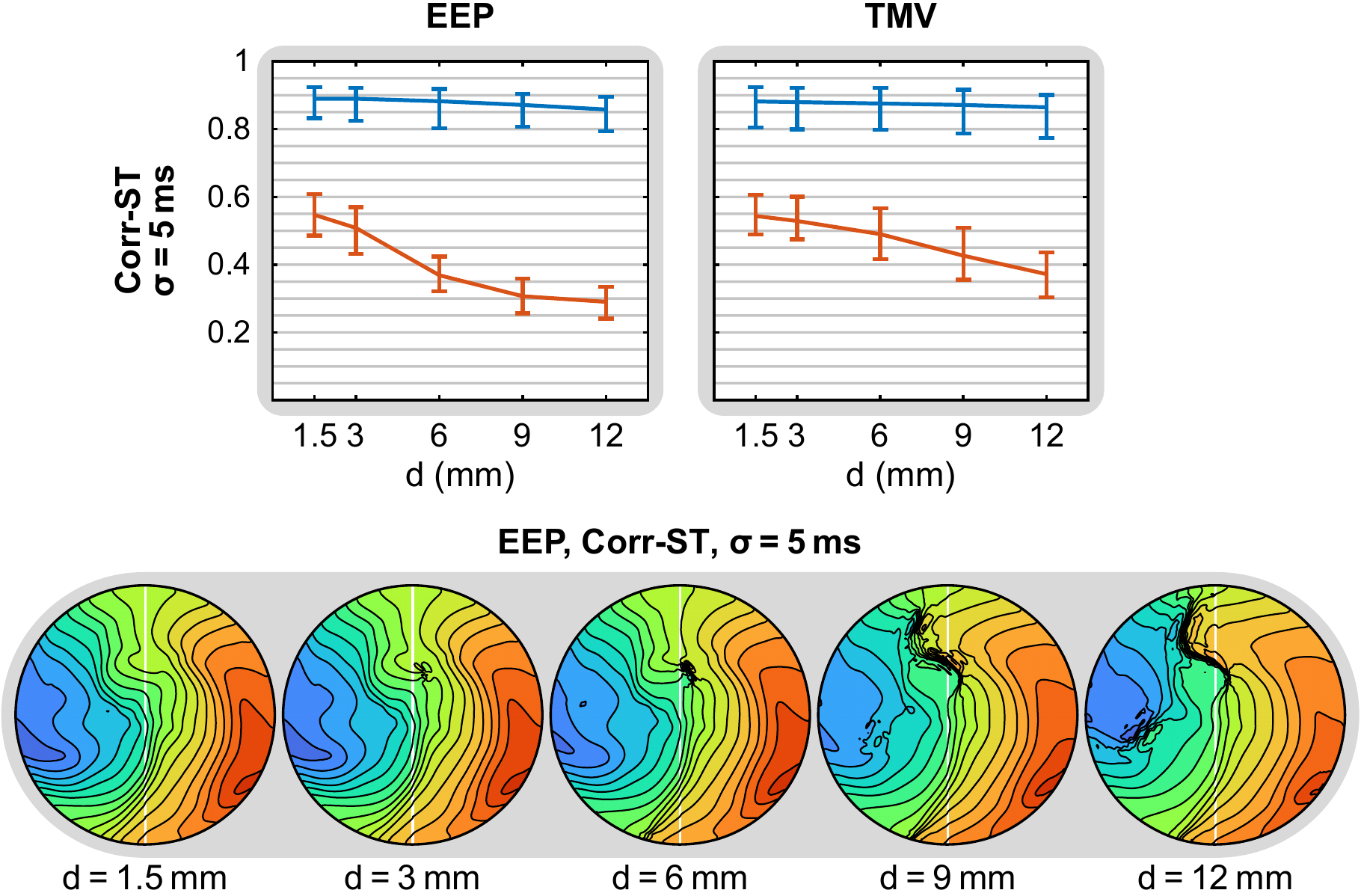}
    \caption{Effect of the node pair distance $d$ used for Corr methods. \textit{Top:} Metrics for Corr-ST on the fine mesh. \textit{Bottom:} Corresponding ATs for EEPs and the same excitation origin as shown in Fig.~\ref{fig:metrics_polar_res}.}
    \label{fig:metrics_polar_d}
    \vspace{-4mm}
\end{figure}

Given its strong impact, we varied $d$ between $1.5$ and $12\,\mathrm{mm}$ for Corr-ST on the fine mesh (Fig.~\ref{fig:metrics_polar_d}). As before, subdivision and Laplacian interpolation was used for $d\,{=}\,1.5\,\mathrm{mm}$. LoB artifacts gradually emerged with increasing $d$ and further small improvements in metrics could be obtained for $d\,{<}\,3\,\mathrm{mm}$.

\subsection{Real Line of Block}
\label{sec:block}
Based on the results in the previous section, the following three configurations are most robust to LoB artifacts:
\begin{itemize}
    \item TMV, Defl-ST, $\sigma\approx 60\,\mathrm{ms}$
    \item TMV, Corr-ST, $\sigma\approx 30\,\mathrm{ms}$, $d\leq 3\,\mathrm{mm}$
    \item EEP, Corr-ST, $\sigma\approx 5\,\mathrm{ms}$, $d\leq 3\,\mathrm{mm}$
\end{itemize}

\begin{figure*}[ht]
    \centering
    \includegraphics[width=0.98\linewidth]{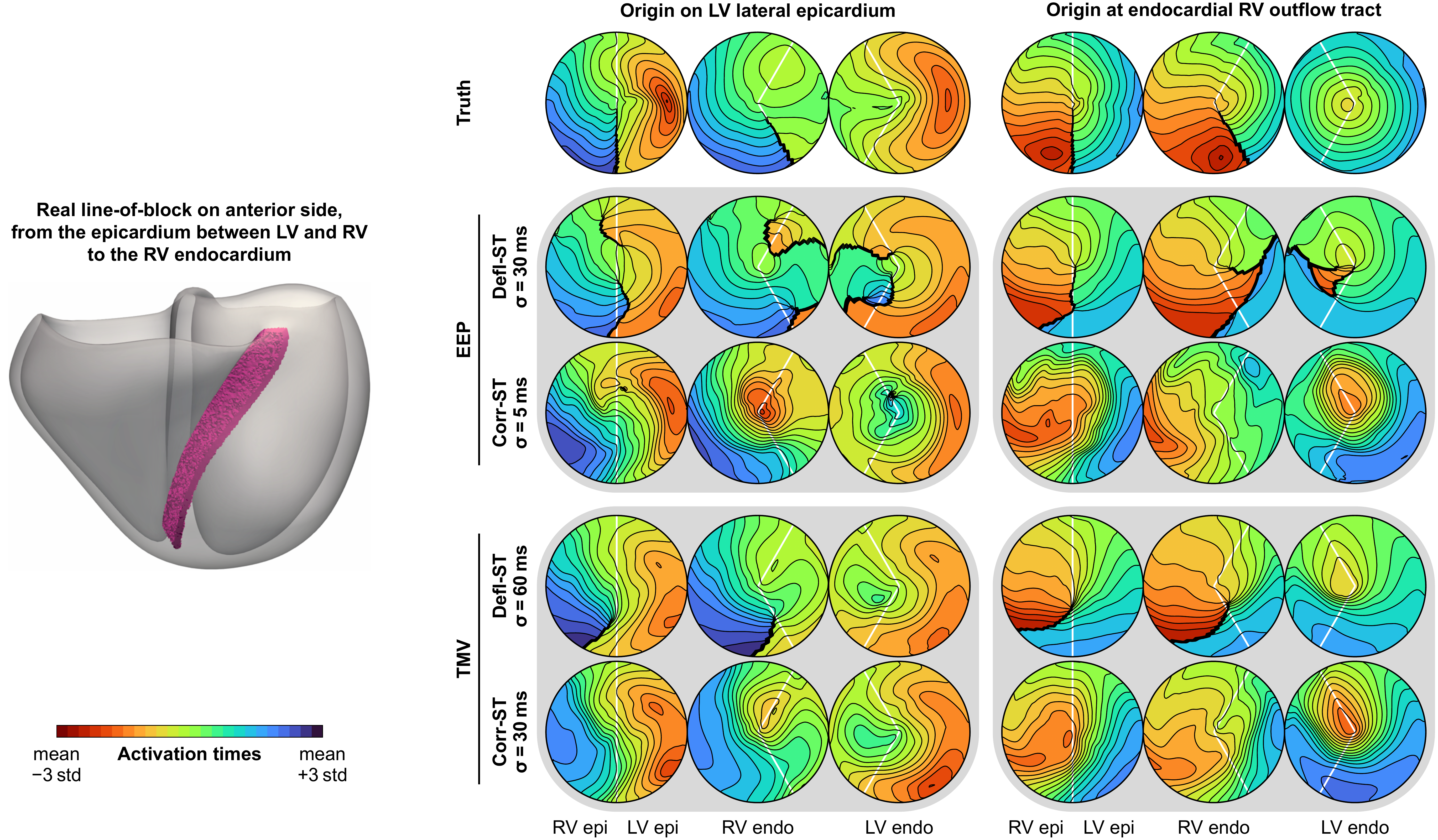}
    \caption{ATs on the epi- and endocardium for two excitation simulations including a real line of block.}
    \label{fig:polar_block_epiEndo}
    \vspace{-4mm}
\end{figure*}

\noindent As robustness to LoB artifacts might be linked to an inherent smoothing effect on ATs, a natural question is whether these configurations are still sensitive to real regions of slow conduction. To this end, we simulated a real LoB for all 137 excitation origins. The LoB cuts off the anterior wall of the RV from the LV and the septum (Fig.~\ref{fig:polar_block_epiEndo}, left) and was modeled as a thin, transmural layer with zero conduction velocity. Although this scenario might not be particularly realistic, it can be considered an extreme case of slow conduction that an ECGI method should be able to detect. The first row of Fig.~\ref{fig:polar_block_epiEndo} (right) depicts the true epi- and endocardial ATs for two exemplary excitation origins resulting from the simulation. Note that the LoB coincides with the radial lines marking the anterior boundary between the septum and the free walls in the polar projections of the epi- and the RV endocardium. Below, the ATs estimated using the three candidate configurations are shown. For completeness, also Defl-ST was included for EEPs. It can be seen that Corr-ST methods failed to recover the real LoB for both source models. The Defl-ST method in combination with TMVs was the only configuration able to detect the real LoB on both the epicardium and the RV endocardium. For EEPs, Defl-ST was also sensitive to the real LoB on the epicardium but not on the RV endocardium. Furthermore, multiple LoB artifacts occurred, making it impossible to discriminate between real and artifactual LoBs.

\vspace{-2mm}
\subsection{Factors Contributing to AT Artifacts}
\label{sec:factorsContrib}

\subsubsection{Signal Morphology and Smoothing}
\label{sec:sigMorphSmoothing}
\ \par
As high spatial frequencies are attenuated by the torso, an effective way to obtain a unique and stable inverse mapping is to suppress these frequencies in the reconstructed sources.
This can be done either explicitly (second-order Tikhonov) or implicitly (zero-order Tikhonov, truncated SVD, GMRES) and explains why many regularization methods lead to spatially smoothed source signals.
Duchateau et al. pointed out that the steepest temporal slope is a high-frequency feature and concluded that smoothing makes it an imprecise AT marker and leads to false AT gradients~\cite{duchateau17}.
However, it is not obvious how spatial smoothing can lead to high-frequency distortions in the spatial arrangement of times of steepest temporal slope, although it is known that spatial smoothing of signals representing the propagation of a depolarization wave also has a ``smoothing'' effect in the temporal domain~\cite[Fig.~3]{schuler19}.
Furthermore, it remains unclear why strong temporal smoothing can help for TMVs but not for EEPs.

\begin{figure*}
    \centering
    \vspace{-4mm}
    \includegraphics[width=0.97\linewidth]{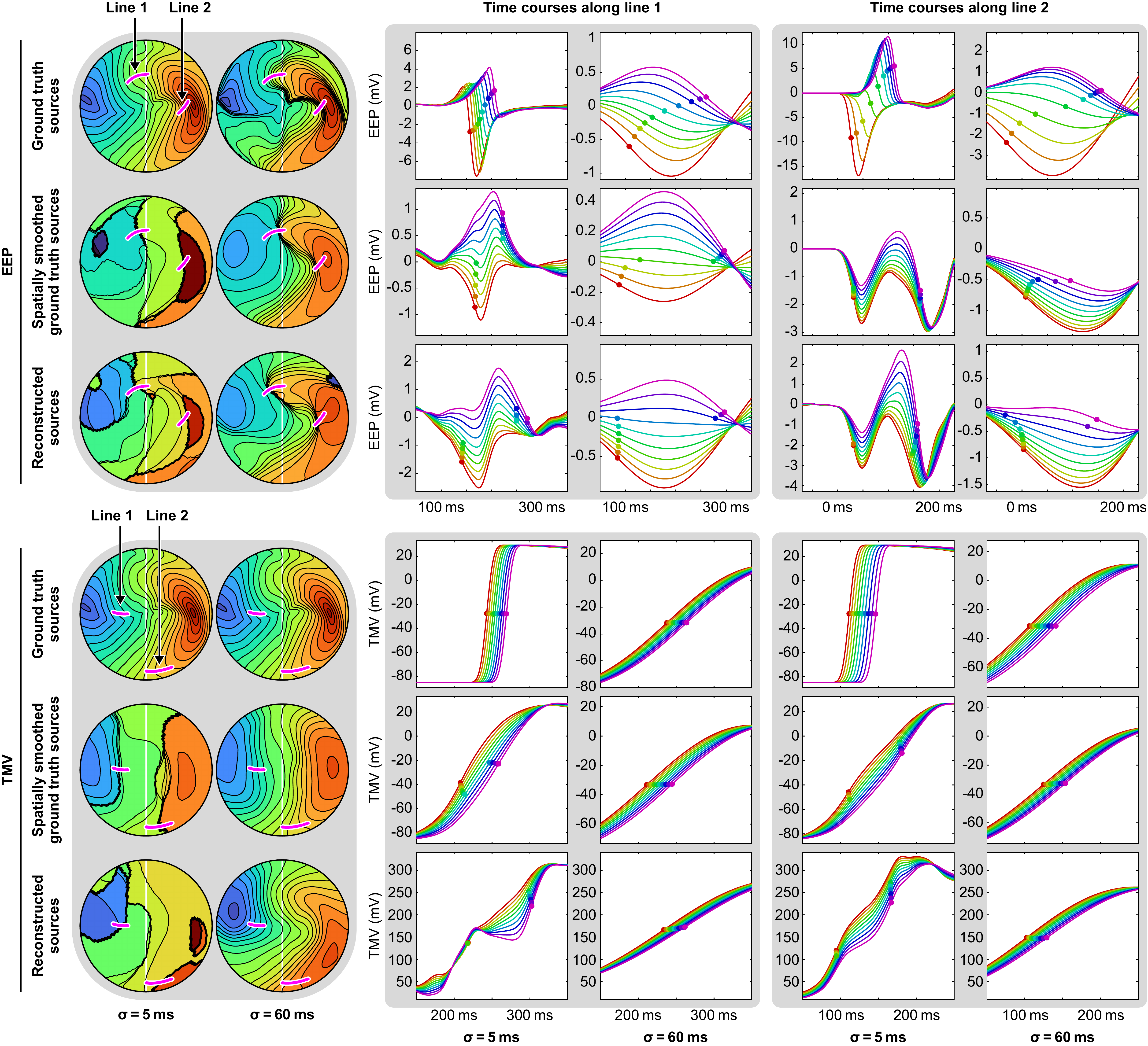}
    \caption{Effect of spatial and temporal smoothing of sources on ATs. \textit{Left:} AT maps estimated using the Defl-T method. See Fig.~\ref{fig:metrics_illustration} for the ground truth and the color bar. \textit{Right:} Time courses of source signals across two LoB artifacts. Dots mark points of steepest down- or upslope.}
    \label{fig:polar_smoothing}
\end{figure*}

To follow up on these questions, we applied spatial and temporal smoothing to the ground truth sources on the fine mesh and observed the related changes in estimated ATs and temporal signal morphologies. Results of this experiment are presented in Fig.~\ref{fig:polar_smoothing} for both EEPs (upper half) and TMVs (lower half).
Each gray panel shows the results for the ground truth sources before spatial smoothing (first row), the ground truth sources after spatial smoothing (second row; see Supplement~S1 for methodological details) and, for comparison, also the inversely reconstructed sources (third row). The two columns represent little and strong temporal smoothing.
The first column in the leftmost panel reveals that spatial smoothing alone is able to reproduce similar LoB artifacts as observed for the reconstructions.
Time courses in the right two panels were extracted from 10 equidistant nodes (distance: $3\,\mathrm{mm}$) on two lines across artifactual LoBs. These lines are marked in magenta in the corresponding AT maps.

\begin{figure*}
    \centering
    \includegraphics[width=\linewidth]{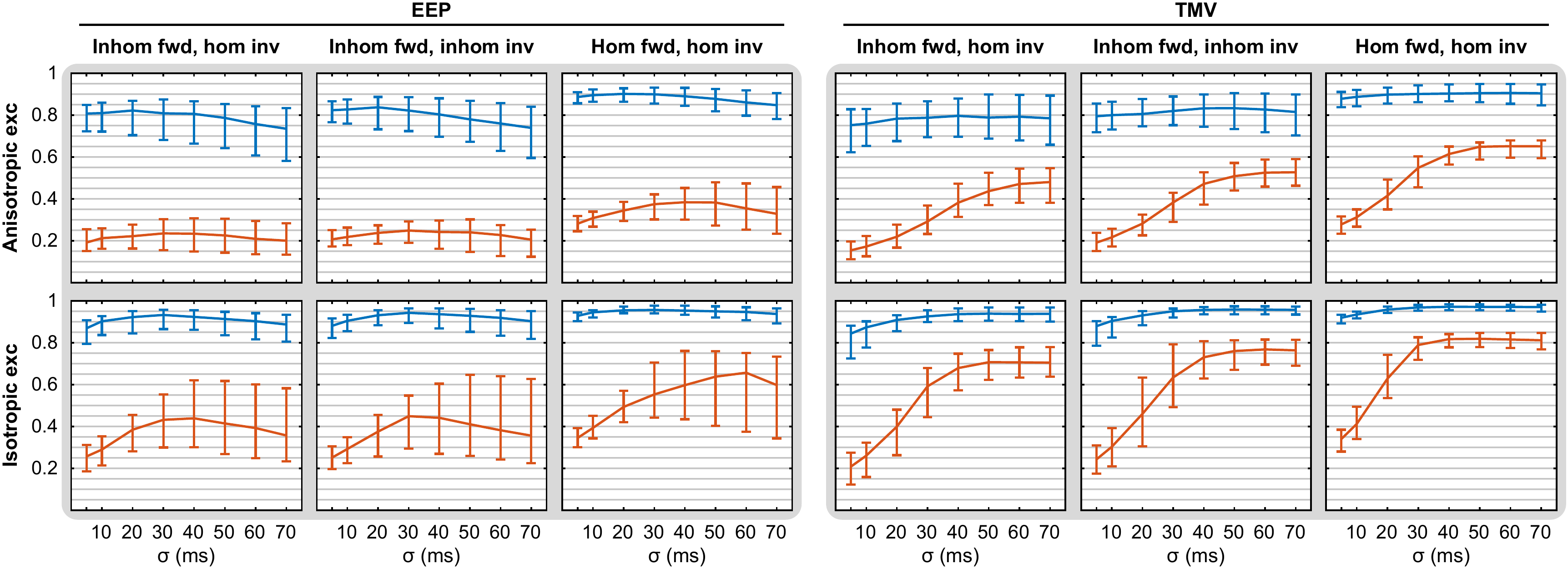}
    \caption{Influence of anisotropy in excitation (exc) simulations and inhomogeneities in forward (fwd) and inverse (inv) calculations on the resulting ATs estimated using the Defl-ST method. See Supplement~S2 for examples of corresponding AT maps and the results with Corr-ST.}
    \label{fig:metrics_aniso_inhom_Defl-ST}
\end{figure*}

For EEPs, the ground truth signals before spatial smoothing show only one downslope. However, their morphology can change from monophasic negative over biphasic to monophasic positive within a range of a few millimeters. Averaging these signals during spatial smoothing can cause additional deflections. For line~1, for example, averaging the red and magenta signals before spatial smoothing (ground truth sources) results in a similar morphology as observed for the green signal after spatial smoothing. For line~2, more surrounding signals not on this line have to be taken into account to explain the resulting morphology. For both lines, the spatially smoothed signals (second row) show strong similarity with the inversely reconstructed signals (third row). In all cases, an additional upslope separating two downslope portions appears and the times of steepest downslope (marked by dots) concentrate in two distinct regions before and after this upslope. This explains the LoB artifacts in AT maps as well as the homogeneous regions on either side of an LoB artifact.\\
Temporal smoothing (second column) cannot resolve all LoB artifacts caused by spatial smoothing and even on its own (first row) can induce LoB artifacts, because the time of steepest downslope only stays at the original position for symmetrically biphasic signals, while it moves left for monophasic negative and right for monophasic positive signals.

For TMVs, the ground truth signals all have the same morphology and therefore spatial smoothing does not create additional deflections. Instead, two inflection points occur after spatial smoothing, which also leads to two distinct regions of steepest slope. However, temporal smoothing can recover one ``global'' inflection point and thereby resolve associated AT artifacts. In contrast to EEPs, temporal smoothing is less problematic for TMVs, because their time course during depolarization is point-symmetric and therefore the inflection point only moves little in time, even for strong smoothing.

\subsubsection{Anisotropic Excitation and Torso Inhomogeneities}
\label{sec:anisoInhom}
\ \par
The anisotropy of excitation propagation and the inhomogeneity of the torso strongly influence BSPs. Although these parameters are determined by the body and cannot be chosen freely, they were varied to assess their effect on AT artifacts. As no perfect forward model can be achieved in practice, the impact of an imperfect model assumed for inverse reconstructions is of particular interest. We computed additional results for isotropic excitation simulations ($0.6\,\mathrm{m/s}$) and three combinations of inhomogeneous (as in Fig.~\ref{fig:torso}) and homogeneous (only heart and rest of the torso) torso models used for forward and inverse calculations.
Fig.~\ref{fig:metrics_aniso_inhom_Defl-ST} shows metrics for ATs estimated using the Defl-ST method. For both source models, the reconstruction was more accurate when the underlying ground truth excitation was isotropic compared to anisotropic excitations (second vs. first row). The detrimental effect of using a homogeneous instead of an inhomogeneous torso model for inverse reconstructions (first vs. second column) was small. However, results were substantially better when a homogeneous torso was not only assumed for the inverse solution but also underlying the ground truth forward calculations (third column). The best metrics resulted for a homogeneous forward and inverse model in combination with isotropic excitation (bottom-right). For EEPs, particularly $r_\mathrm{SN}$ increased for both an isotropic excitation and a homogeneous forward model, indicating that both anisotropic excitation and torso inhomogeneities contribute to AT artifacts.

\section{Discussion}
\label{sec:discussion}

\subsection{Line-of-block Artifacts}
\label{sec:lobArtifacts}
We were able to reproduce and analyze LoB artifacts reported in clinical and experimental studies \cite{duchateau18,bear19} using computer simulations. Our main findings are:
\begin{itemize}
\item LoB artifacts are mainly caused by the spatial smoothing effect of the regularization used to solve the ill-posed inverse problem, which leads to additional deflections or inflections in the time course of reconstructed signals.
\item For TMVs, sufficient temporal smoothing of reconstructed signals can eliminate LoB artifacts by recovering one global slope of deflection.
\item For EPs, temporal smoothing cannot eliminate LoB artifacts because, in contrast to TMVs, the morphology of their time course varies substantially in space.
\end{itemize}

The morphology of EPs depends on several factors.
One factor is the drift of the reference potential from the positive to the negative extreme of the potential distribution during the spread of excitation~\cite{taccardi98}. It has been suggested to remove this drift~\cite{punske99,ghodrati06}, but it is unclear whether such a removal can reliably be applied to reconstructed EPs.
Furthermore, EPs are a secondary phenomenon resulting from a spatial superposition of local and distant activity. This leads to a dependence on the curvature and transmurality of the wavefront.
Other factors affecting the morphology of EPs are anisotropic excitation and torso inhomogeneities. This could explain the improvements in $r_\mathrm{SN}$ observed for a homogeneous forward calculation and an isotropic excitation in Fig.~\ref{fig:metrics_aniso_inhom_Defl-ST}.

TMVs do not depend on the reference potential and -- in theory -- only represent local activity that follows a stereotypical time course. Consequently, TMVs share a uniform morphology. However, a spatially constant offset remains undetermined in reconstructed TMVs, requiring a baseline correction~\cite{schuler19a}. This baseline correction in conjunction with subsequent temporal smoothing exploits the temporal structure of TMVs to overcome ambiguities during AT estimation (only one deflection remains after sufficient smoothing) and may be seen as a way to leverage prior knowledge for AT estimation.

Although outside the scope of this study, it is worth noting that an alternative approach with even stronger assumptions on the morphology of sources is direct, non-linear AT imaging~\cite{huiskamp97} using a temporal parameterization of TMVs (EDL sources). While promising results have been reported~\cite{dam09,janssen18}, this approach is sensitive to the initialization of the non-convex optimization and to cases where TMVs may not match the assumed waveform, such as in the case of depression of TMV magnitude in ischemia~\cite{erem14a}.
In linear formulations of the inverse problem, on the other hand, spatial smoothing of reconstructed sources may be mitigated as well by using temporal priors for inverse reconstructions~\cite{messnarz04,yu15,schuler19a}.

\subsection{AT Estimation Methods}
\label{sec:atEstimMethods}
The most important observations from the comparison of AT estimation methods are:
\begin{itemize}
\item Methods using the spatiotemporal derivative perform better than methods based purely on the temporal derivative.
\item Correlation-based (Corr) methods (with a small node pair distance) are more robust to LoB artifacts than deflection-based (Defl) methods but do not preserve real LoBs.
\item For TMVs, similar robustness as with Corr methods can be reached with Defl methods by increasing the temporal smoothing, while retaining the sensitivity to real LoBs.
\end{itemize}

Using the spatiotemporal derivative may be seen as exploiting prior knowledge valid for both EPs and TMVs because both show a gradient in space and time during activation.

Regarding the capability of Corr methods to preserve real LoBs, we also tested a least-absolute-deviation instead of a least-squares regression and weights based on the correlation coefficient as in~\cite{duchateau17} but neither led to a notable improvement.

\subsection{Source Models}
\label{sec:sourceModels}
Apart from the differences regarding LoB artifacts, we observed that TMVs lead to more accurate ATs on the septum than EEPs (Fig.~\ref{fig:polar_sm_epiEndo} and Fig.~\ref{fig:metrics_sm}, right). This can be explained by visualizing the magnitude of transfer coefficients (TC) on the heart surface (see Supplement~S3 for methodological details). Fig.~\ref{fig:polar_transferCoeffs} shows that the TCs for EEPs are much smaller on the endo- than on the epicardium. This means that endocardial EPs contribute only little to BSPs and that the contribution of endocardial activity to BSPs is already largely contained in epicardial EPs. As a consequence, it is very difficult to delineate endo- and epicardial activity in reconstructed EPs. Transmural regularization helps to determine EPs on the non-septal endocardium. However, as TCs are small on both sides of the septal endocardium, EPs on the septum are essentially just a smooth extrapolation from non-septal regions. The small endocardial TCs also explain why on the epicardium, PPs, which were reconstructed using a non-transmural Laplacian, were very similar to EEPs reconstructed using a transmural Laplacian.
For TMVs, TCs have a similar magnitude on the epi- and endocardium. In principle, this allows reconstructing also endocardial activity.
However, transmural regularization is essential because TCs on opposite sides of the myocardial wall show strong linear dependence (opposite signs). Furthermore, ambiguities exist between the septum and the free walls, which may only be resolved using temporal regularization~\cite{messnarz04,schuler19a}.

\begin{figure}[ht]
    \centering
    \includegraphics[width=0.98\linewidth]{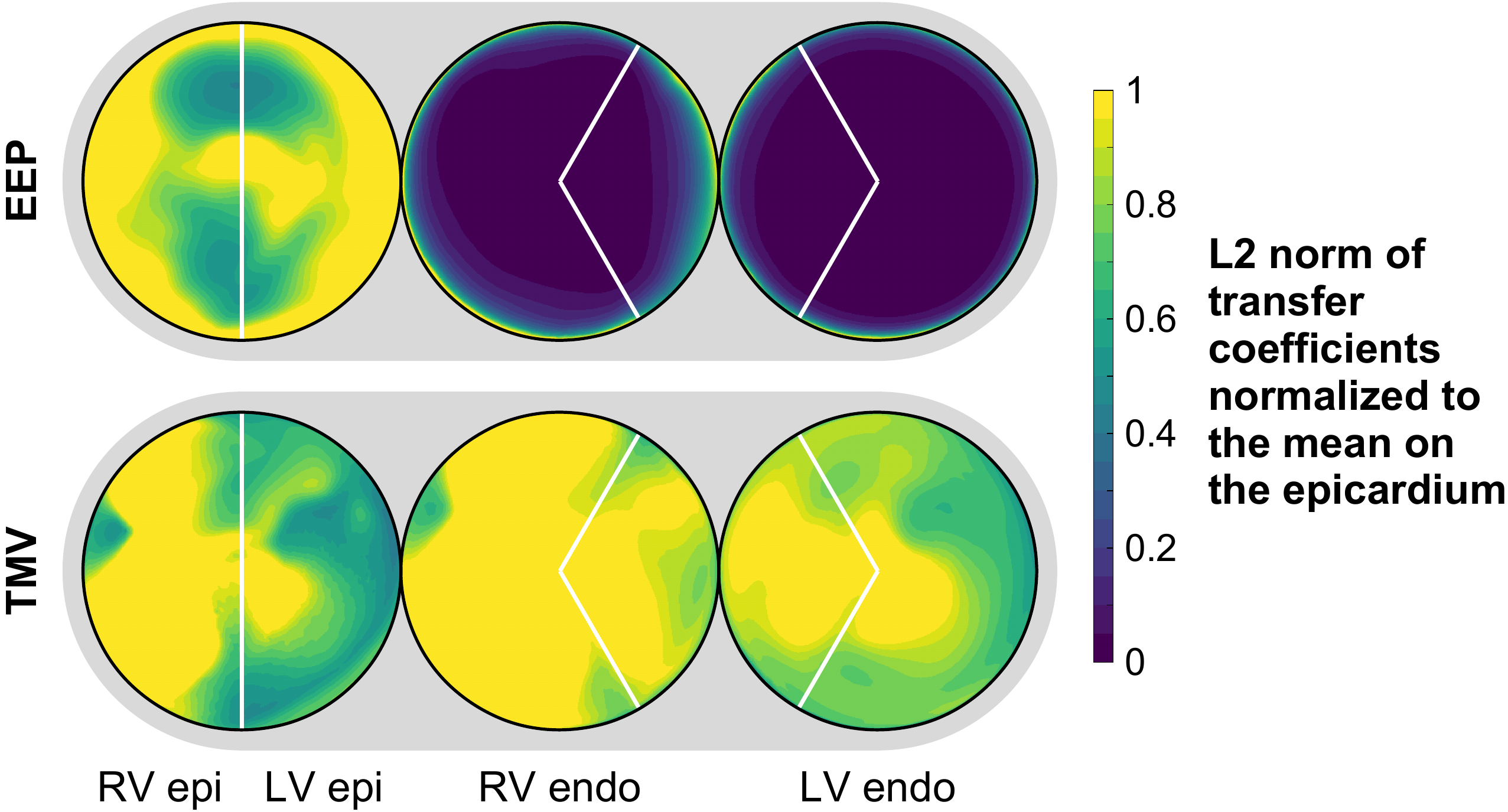}
    \caption{Transfer coefficients for EEPs show a much smaller magnitude on the endo- than on the epicardium. This is not the case for TMVs.}
    \label{fig:polar_transferCoeffs}
\end{figure}
\vspace{-3mm}

\subsection{Mesh Resolution}
\label{sec:meshResolution}
The impact of mesh resolution used for inverse reconstructions and for computation of the spatial derivative turned out to be negligible, at least as long as it is no larger than $12\,\mathrm{mm}$, the largest value tested here. However, a small node pair distance \mbox{(${<}\,3\,\mathrm{mm}$)} is important for Corr methods to guarantee sufficient morphological similarity of cross-correlated signals. A coarse mesh may be used for reconstructions, but then sources should be interpolated onto a finer mesh to allow estimation of time delays across shorter distances. These findings only apply to regularization methods that yield spatially smooth sources.

\subsection{Limitations and Outlook}
\label{sec:limitations}
In this work we assumed isotropic myocardial conductivities. This is a typical simplification for solving the inverse problem~\cite{janssen18}. Using the same assumption also for forward calculations allowed us to exclude a mismatch of torso models used in forward and inverse calculations as a potential cause of AT artifacts. However, this assumption does not reflect reality~\cite{potse09} and future work should target the impact of anisotropic myocardial conductivities on estimated ATs.

Furthermore, only white noise was assumed, whereas clinical recordings also include correlated noise. Correlated noise was not required to reproduce LoB artifacts, but its effect on spatially (and temporally) smoothed inverse solutions may further complicate AT estimation and may be worth studying.

Investigating ways to remove the reference drift~\cite{taccardi98} from EPs was out of the scope of this study. Given its influence on the temporal morphology of EPs and thus on AT estimation, this might be an important aspect for future research.

Similar investigations may improve the estimation of repolarization times using ECGI, as suggested in~\cite{cluitmans17,bear21}.
In this case, the action potential duration and shape should be varied.

\section{Conclusion}
\label{sec:conclusion}
LoB artifacts are mainly caused by spatial smoothing during regularization. Differences between EPs and TMVs can be explained by their temporal morphology and their transfer coefficients. Although Corr-based methods for AT estimation are generally less prone to LoB artifacts, only Defl methods in combination with TMVs and strong temporal smoothing can prevent LoBs artifacts while also preserving real LoBs.

\bibliographystyle{ieeetran}
\bibliography{references}

\clearpage
\onecolumn
\setcounter{section}{0}
\setcounter{figure}{0}
\renewcommand{\thesection}{S\arabic{section}}
\renewcommand{\thefigure}{S\arabic{figure}}
\renewcommand{\thetable}{S\arabic{table}}

{
\centering\sf
{\color{nblue}\Large
Supplementary Materials:\\\vspace{0.15mm}
Reducing Line-of-block Artifacts in Cardiac Activation Maps Estimated Using\\\vspace{0.15mm}
ECG~Imaging: A Comparison of Source Models and Estimation Methods\\}
\vspace{3mm}
{\normalsize
Steffen Schuler*, Matthias Schaufelberger, Laura R. Bear, Jake A. Bergquist, Matthijs J. M. Cluitmans,\\ Jaume Coll-Font, Önder N. Onak, Brian Zenger, Axel Loewe, Rob S. MacLeod, Dana H. Brooks, and Olaf Dössel}
}
\vspace{3mm}

\section{Spatial Smoothing of Ground Truth Sources}
\label{sec:appSpatialSmoothing}
Spatial smoothing of the ground truth sources in section~III-D.1 was performed as:
\begin{equation*}
    \mathbf{x}_\mathrm{smooth} = \arg\min_{\hat{\mathbf{x}}}\bigl( \|\hat{\mathbf{x}}-\mathbf{x}\|_2^2 + \mu\,\|\mathbf{L}\mathbf{x}\|_2^2 \bigr) = \underbrace{(\mathbf{I}+\mu\,\mathbf{L}^\top\mathbf{L})^{-1}}_\mathbf{M}\,\mathbf{x}
\end{equation*}
$\mathbf{x}$ are the ground truth sources for one time step and $\mathbf{M}$ is the spatial smoothing operator.
For smoothing of TMVs, we used the same transmural Laplacian as for the inverse reconstructions. For smoothing of EEPs, however, we used a non-transmural Laplacian, because the magnitude of transfer coefficients for EEPs is small on the endocardium compared to the epicardium (Fig.~15) and therefore using a non-transmural Laplacian more closely resembles the smoothing effect of the reconstruction on epicardial EEPs. $\mu$ was chosen so that the half-width at half-maximum of the average point spread function of $\mathbf{M}$ was $30\,\mathrm{mm}$ for EEPs and $40\,\mathrm{mm}$ for TMVs, which resulted in similarly strong smoothing as observed for reconstructions.

\section{Impact of Anisotropic Excitation and Torso Inhomogeneities}
\label{sec:appAnisoInhom}
\begin{figure*}[ht]
    \centering
    \includegraphics[width=0.75\linewidth]{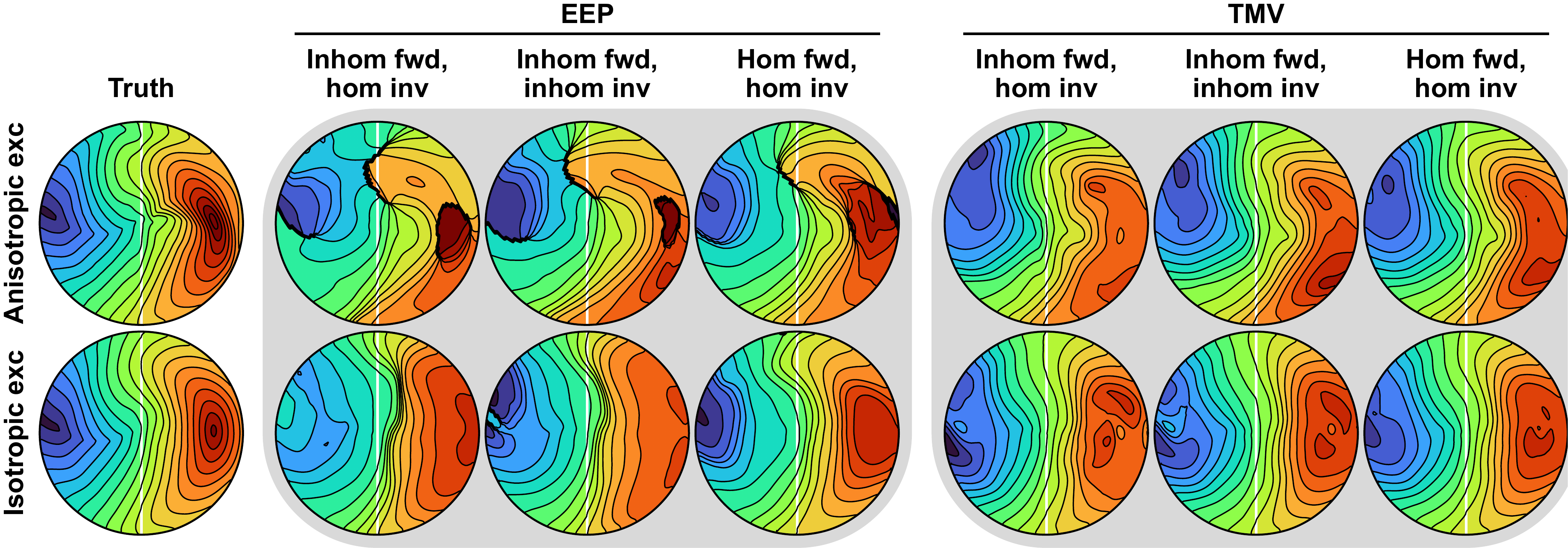}
    \caption{Influence of anisotropic excitation (exc) and torso inhomogeneities in forward (fwd) and inverse (inv) calculations on the resulting AT maps estimated using the Defl-ST method for EEPs ($\sigma\,{=}\,30\,\mathrm{ms}$) and TMVs ($\sigma\,{=}\,60\,\mathrm{ms}$). See Fig.~6 for a color bar.}
    \label{fig:polar_aniso_inhom_Defl-ST}
\end{figure*}

\begin{figure*}[ht]
    \centering
    \includegraphics[width=\linewidth]{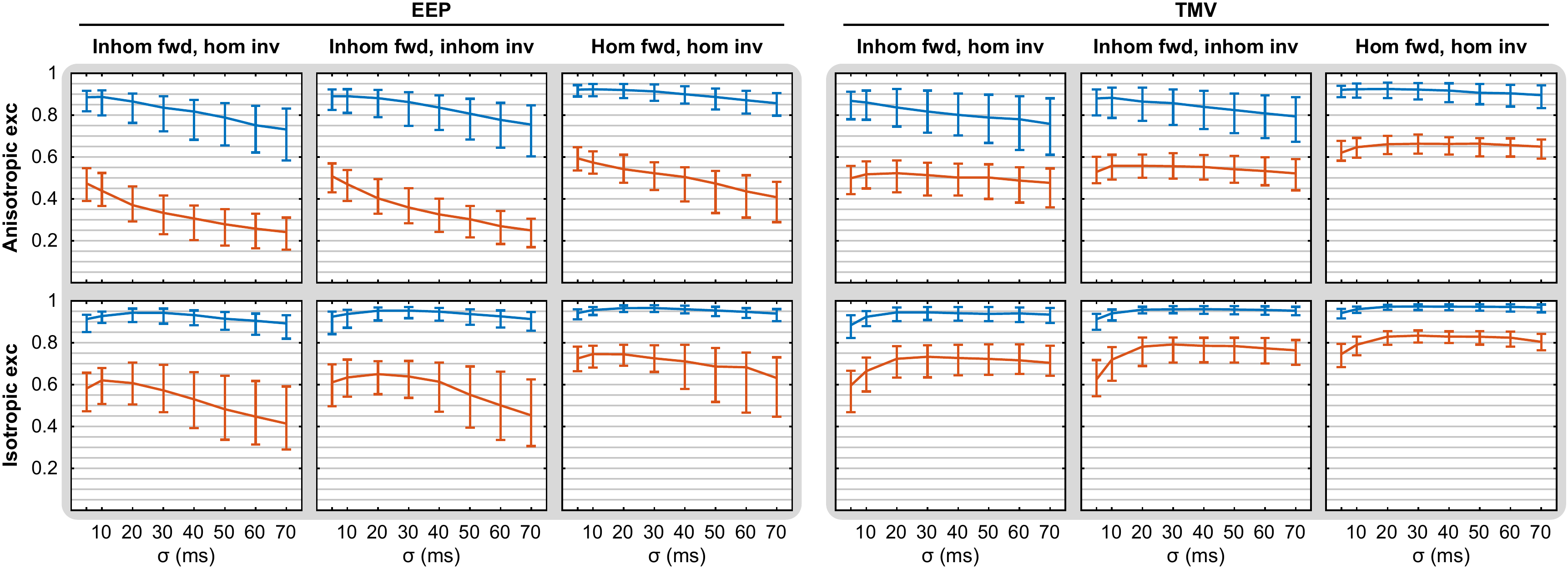}
    \caption{As Fig.~14 but for the Corr-ST method.}
    \label{fig:metrics_aniso_inhom_Corr-ST}
\end{figure*}

Fig.~\ref{fig:polar_aniso_inhom_Defl-ST} depicts the impact of anisotropic excitation and torso inhomogeneities on exemplary AT maps estimated using the Defl-ST method.
Fig.~\ref{fig:metrics_aniso_inhom_Corr-ST} shows AT metrics corresponding to Fig.~14 but for the Corr-ST instead of the Defl-ST method.
\vspace{-3mm}

\newpage
\section{Visualizing the Magnitude of Transfer Coefficients}
\label{sec:appTransferCoeffs}
The magnitude of transfer coefficients depicted in Fig.~15 was computed as $\|\mathbf{A}\,\mathbf{M}^{-1}\|_\mathrm{col}$.
Here, $\mathbf{M}$ is the mass matrix, $\mathbf{A}$ the transfer matrix and $\|\cdot\|_\mathrm{col}$ denotes the $L^2$ norm along columns (across electrodes). Visualizing the result on the surface of the heart has some similarity with the contribution maps introduced by van Oosterom and Huiskamp\footnote{A. van Oosterom and G. J. Huiskamp, ``The effect of torso inhomogeneities on body surface potentials quantified using ``tailored'' geometry,'' J.~Electrocardiol., vol. 22, pp. 53–72, 1989.}, but instead of the contribution of a point on the heart to a single electrode on the body surface, it shows the strength by which a point contributes to all $M$ electrodes (as measured by the length of the $M$-dimensional vector representing the corresponding BSP pattern).

The small transfer coefficients for EEPs on the endocardium correspond to the small sensitivity of solutions of Laplace's equation outside the heart's convex hull to Dirichlet boundary conditions on concave parts of the heart surface.

\section{Impact of the Regularization Parameter}
\label{sec:appReguParam}
To make sure that the differences in source models are not due to the regularization parameters, we computed additional results for ``optimal'' parameters that maximize the correlation between the gradient norms of reconstructed and true sources over a grid of 100 log-spaced values of $\lambda$ across 8 decades:
\begin{equation*}
\lambda_\mathrm{optimal} = \arg\underset{\lambda}{\max}\Bigl\{\underset{\forall i,t}{\operatorname{corr}}\left(\|\nabla \widehat{x}_{i,t}(\lambda)\|_2, \|\nabla x_{i,t}^\mathrm{truth}\|_2\right)\Bigr\}
\end{equation*}
$\widehat{x}_{i,t}$ and $x_{i,t}^\mathrm{truth}$ are the reconstructed and true sources at node $i$ and time step $t$, respectively. $\nabla$ denotes the surface gradient and $\operatorname{corr}$ the Pearson correlation computed over space and time (only the true activation interval was taken into account).

\begin{figure}[ht]
    \vspace{-1.5mm}
    \centering
    \includegraphics[width=0.35\linewidth]{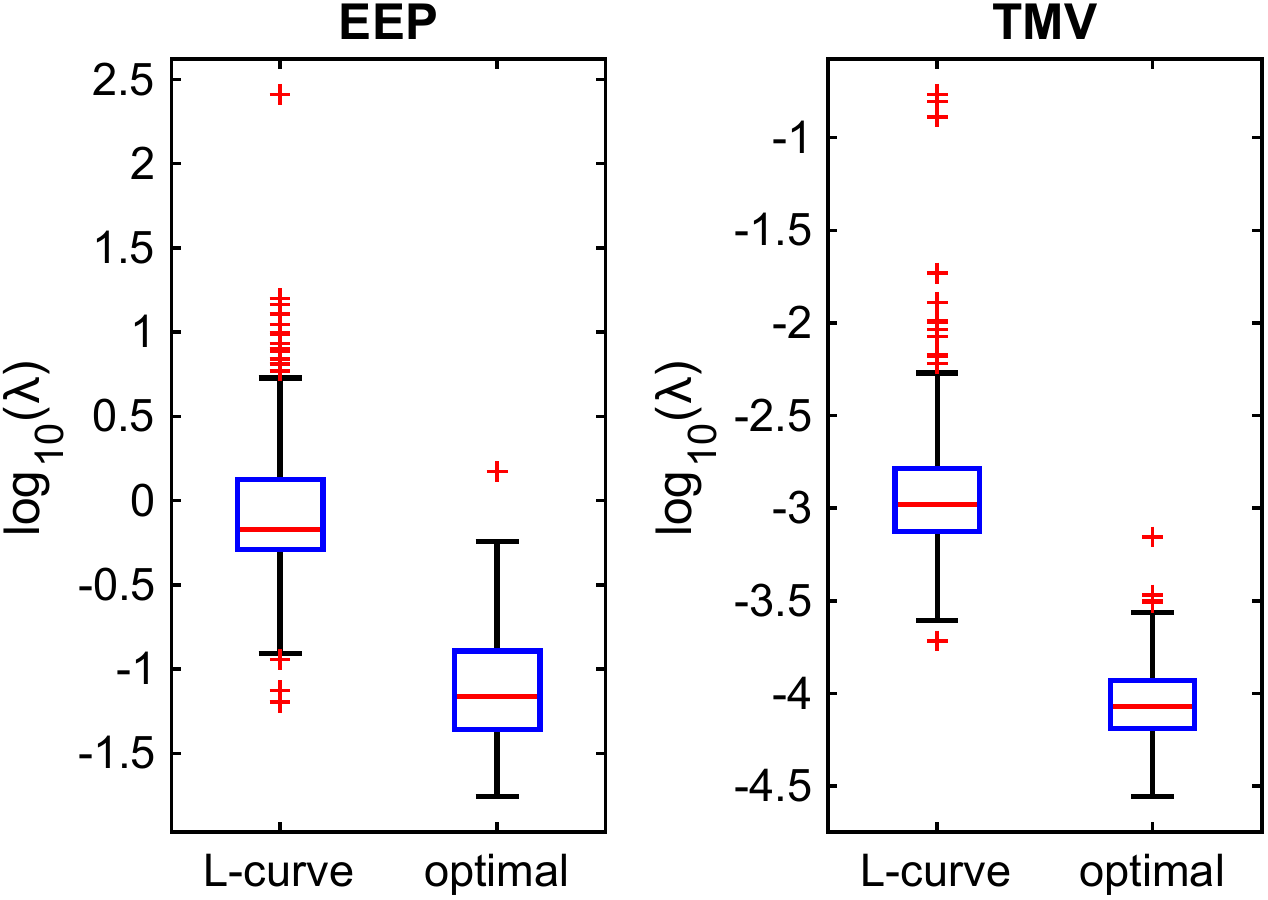}
    \qquad\qquad
    \includegraphics[width=0.35\linewidth]{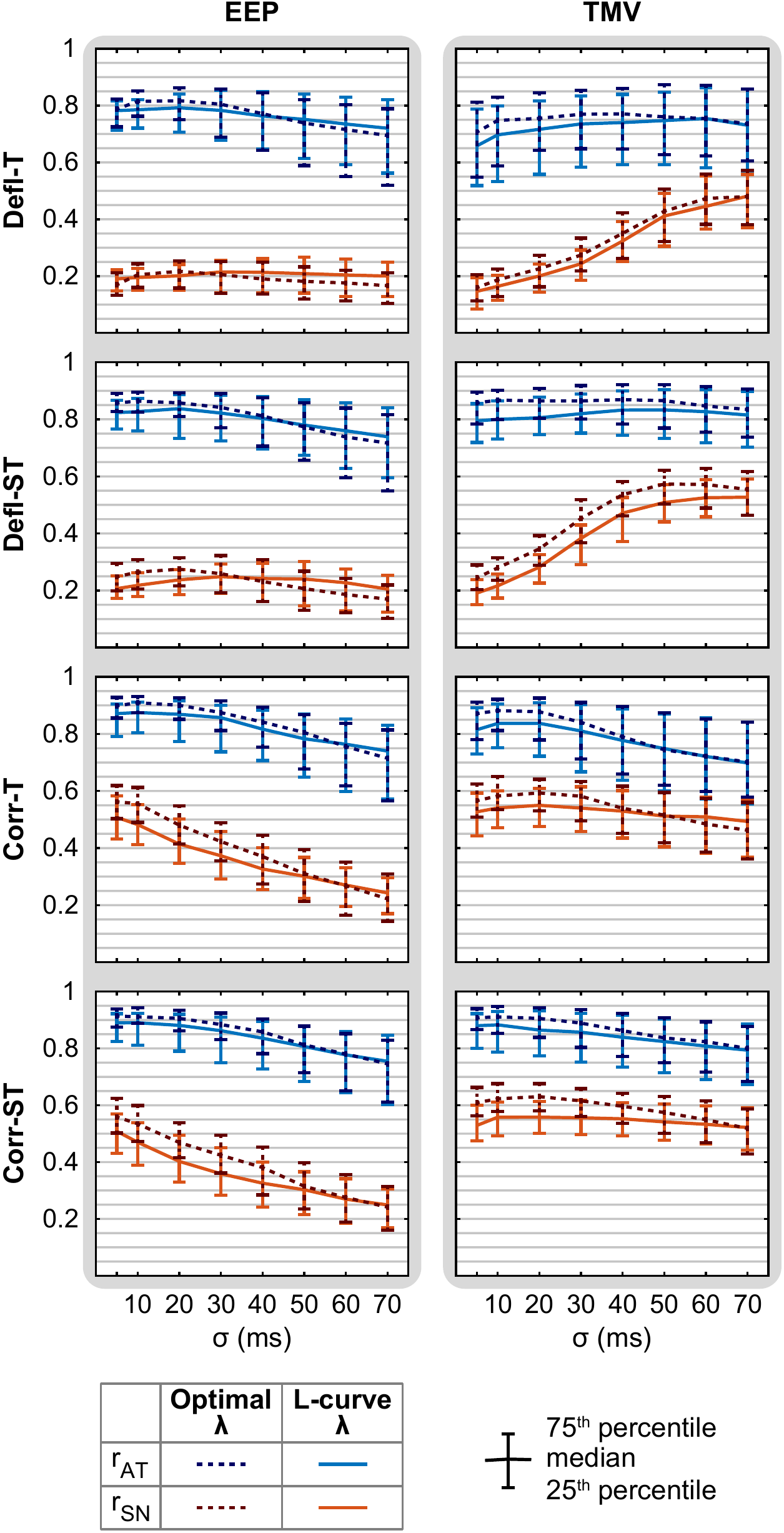}
    \caption{\textit{Left:} Regularization parameters determined using the L-curve criterion (L-curve~$\lambda$) and optimized using the ground truth sources (optimal~$\lambda$). Each box plot contains all 137 excitation origins. \textit{Right:} Resulting AT metrics for the optimal $\lambda$ in comparison to the L-curve $\lambda$.}
    \label{fig:lambda}
    \vspace{-1.5mm}
\end{figure}

Fig.~\ref{fig:lambda}~(left) reveals that for both source models, the L-curve criterion led to slight overregularization. However, the same trend in AT metrics can be observed for both sets of parameters in Fig.~\ref{fig:lambda}~(right).

\end{document}